\documentclass[conference]{IEEEtran}
\usepackage[available,functional,reproduced]{ndssbadges}
\IEEEoverridecommandlockouts

\usepackage[ruled,vlined]{algorithm2e}
\usepackage{mathtools}
\usepackage{xcolor}
\usepackage{caption}
\usepackage{listings}
\usepackage[T1]{fontenc}
\usepackage[utf8]{inputenc}
\usepackage{colortbl}
\usepackage{etoolbox}
\usepackage{siunitx}
  \robustify\bfseries
  \robustify\bf
\usepackage{xparse}
\usepackage{xspace}

\usepackage[many]{tcolorbox}% for COLORED BOXES (tikz and xcolor included)
 
\usepackage{amsmath, amssymb, amsfonts}
\usepackage{multirow}

\usepackage{xspace}
\usepackage{adjustbox}
\usepackage{graphicx}
\usepackage{pdflscape}
\usepackage{tabularx, booktabs}
\usepackage{makecell}
\usepackage{pifont} % used to get \ding command
\usepackage{wasysym} % more symbols, like circle, rightcircle,..
\usepackage{csquotes}

\usepackage{epigraph}
\usepackage[hyphens]{url}
\usepackage{hyperref}
\hypersetup{breaklinks,colorlinks,anchorcolor=black,citecolor=black,filecolor=black,linkcolor=black,menucolor=black,urlcolor=black}
\usepackage{cite}  
\usepackage{algorithmic}
\usepackage{textcomp}
\usepackage{bmpsize}
\usepackage{lipsum}

\usepackage{pgfplots}
\pgfplotsset{compat=1.18}
\usepackage{array}
\usepackage{pgf-pie}
\usetikzlibrary{patterns}
\usetikzlibrary{calc, matrix, patterns}
\usetikzlibrary{positioning}
\pgfdeclarelayer{background}
\pgfsetlayers{background,main}
\usetikzlibrary{fit}
\usetikzlibrary{matrix}
\usepgfplotslibrary{fillbetween}
\usepackage{pgfplotstable}

\usepackage{wrapfig}
\usepackage{standalone}

\renewcommand{\paragraph}[1]{{\vskip 5pt \noindent\textit{#1.}}}

\newcommand\dataset[1]{\emph{#1}}  % use this to wrap dataset names in a common style
\newcommand\entity[1]{#1}  % use this to wrap entities like Huggingface, Github, OpenAI, Google etc. in a common style
\newcommand\model[1]{#1}  % use this to wrap model name like Gemini, GPT-3, ChatGPT~3.5~Turbo, etc. in a common style
\newcommand\modelID[1]{\emph{#1}}  % use this to wrap model IDs like Phi~3~mini~4k etc. in a common style
\newcommand\cveID[1]{\emph{#1}}  % use this to wrap CVE IDs in a common style

\newcommand{\GPTapp}{ChatGPT\xspace}  % use this to refer to the smartphone app or web app
\newcommand{\GPTmodel}{GPT\xspace}  % use this to refer to the model or the API-Interface

\newcommand{\GPTfour}{GPT-4\xspace}  % https://openai.com/index/gpt-4-research/
\newcommand{\GPTfourturbo}{GPT-4 Turbo\xspace}  % https://platform.openai.com/docs/models/gpt-4-turbo
\newcommand{\GPTfouro}{GPT-4o\xspace}  % https://openai.com/index/hello-gpt-4o/
\newcommand{\GPTthreefiveturbo}{GPT-3.5 Turbo\xspace}  % mhttps://platform.openai.com/docs/models/gpt-3.5-turbo

\newcommand{\eg}{e.\,g.\xspace}
\newcommand{\ie}{i.\,e.\xspace}
\newcommand{\etal}{et~al.\@\xspace}
\newcommand{\cf}{cf.\@\xspace}

\newtcolorbox{promptbox}{
    sharpish corners, % better drop shadow
    colback = nord_gray, %background color
    boxrule = 0pt,
    toprule = 2pt, % top rule weight
    enhanced,
    fuzzy shadow = {0pt}{-2pt}{-0.5pt}{0.5pt}{nord_gray}, % {xshift}{yshift}{offset}{step}{options} 
    width = .9\linewidth
}

\newcommand{\result}[1]{%
\begin{tcolorbox}[colframe=black,boxrule=0.5pt,arc=4pt,
      left=6pt,right=6pt,top=6pt,bottom=6pt,boxsep=0pt,width=\columnwidth]%
      {\emph{#1}}
\end{tcolorbox}%
}

\SetKwComment{Comment}{/* }{ */}

\newcolumntype{R}[2]{%
    >{\adjustbox{angle=#1,lap=\width-(#2)}\bgroup}%
    c%
    <{\egroup}%
}

\newcommand{\boldpar}[1]{\medskip\noindent\textbf{#1.}\xspace}
\newcommand{\italicpar}[1]{\medskip\noindent\emph{#1.}\xspace}

\usepackage{pifont} % used to get \ding command
\usepackage{wasysym} % more symbols, like circle, rightcircle,...

\usepackage{enumitem}
\setlist[itemize]{itemsep=0.5em}

\definecolor{cbone}    {HTML}{006BA4} % blue
\definecolor{cbtwo}    {HTML}{FF800E} % orange
\definecolor{cbthree}  {HTML}{ABABAB} % grey
\definecolor{cbfour}   {HTML}{595959} % dark grey
\definecolor{cbfive}   {HTML}{5F9ED1} % light blue
\definecolor{cbsix}    {HTML}{C85200} % dark orange
\definecolor{cbseven}  {HTML}{898989} % darker grey
\definecolor{cbeight}  {HTML}{A2C8EC} % lighter blue
\definecolor{cbnine}   {HTML}{FFBC79} % light orange
\definecolor{cbten}    {HTML}{CFCFCF} % lighter grey
\definecolor{cbeleven}    {HTML}{A9373B} % dark red

\colorlet{primarycolor}{cbone}
\colorlet{primaryshaded}{cbfive}
\colorlet{secondarycolor}{cbtwo}
\colorlet{secondaryshaded}{cbnine}
\colorlet{secondarymore}{cbsix}
\colorlet{secondaryevenmore}{cbeleven}
\colorlet{tertiarycolor}{cbfour}
\colorlet{tertiaryshaded}{cbten}

\iffalse
\definecolor{colorNotPresent}{HTML}{2369BD}
\definecolor{colorDoesNotApply}{HTML}{678BBE}

\definecolor{colorLikelyPresentBD}{HTML}{9BAECB}
\definecolor{colorLikelyPresent}{HTML}{CFD4E0}

\definecolor{colorUnclearFromText}{HTML}{D3D3D3}

\definecolor{colorPartlyPresentBD}{HTML}{E7C8C6}
\definecolor{colorPartlyPresent}{HTML}{D39794}

\definecolor{colorPresentBD}{HTML}{BF6765}
\definecolor{colorPresent}{HTML}{A9373B}

\else

\colorlet{colorNotPresent}{primarycolor}
\colorlet{colorDoesNotApply}{primaryshaded}

\colorlet{colorLikelyPresentBD}{secondaryshaded}
\colorlet{colorLikelyPresent}{secondaryshaded}

\colorlet{colorUnclearFromText}{cbthree}

\colorlet{colorPartlyPresentBD}{secondarycolor}
\colorlet{colorPartlyPresent}{secondarycolor}

\colorlet{colorPresentBD}{secondarymore}
\colorlet{colorPresent}{secondarymore}%{secondaryevenmore}

\fi

\colorlet{colorSecurityInLLM}{colorNotPresent}
\colorlet{colorLLMForSecurity}{colorDoesNotApply}

\newcommand{\presenceGrid}[9]{%
  \pgfmathtruncatemacro{\nPresent}{#1}%
  \pgfmathtruncatemacro{\nPresentButDiscussed}{#2}%
  \pgfmathtruncatemacro{\nTotalPresent}{#1+#2}%
  \pgfmathtruncatemacro{\nPartlyPresent}{#3}%
  \pgfmathtruncatemacro{\nPartlyPresentButDiscussed}{#4}%
  \pgfmathtruncatemacro{\nLikelyPresent}{#5}%
  \pgfmathtruncatemacro{\nLikelyPresentButDiscussed}{#6}%
  \pgfmathtruncatemacro{\nDoesNotApply}{#7}%
  \pgfmathtruncatemacro{\nNotPresent}{#9}%
  \pgfmathtruncatemacro{\nUnclearFromText}{#8}%
  \def\rectW{2.87}% mm
  \def\rectH{3.45}% mm
  \def\hsep{0.8}% mm
  \def\vsep{0.8}% mm
  \def\rows{3}%
  \def\cols{24}% rows × cols must stay 72 (# papers)
  \pgfmathtruncatemacro{\tA}{\nPresent}%
  \pgfmathtruncatemacro{\tB}{\tA+\nPresentButDiscussed}%
  \pgfmathtruncatemacro{\tC}{\tB+\nPartlyPresent}%
  \pgfmathtruncatemacro{\tD}{\tC+\nPartlyPresentButDiscussed}%
  \pgfmathtruncatemacro{\tE}{\tD+\nLikelyPresent}%
  \pgfmathtruncatemacro{\tF}{\tE+\nLikelyPresentButDiscussed}%
  \pgfmathtruncatemacro{\tG}{\tF+\nUnclearFromText}%
  \pgfmathtruncatemacro{\tH}{\tG+\nDoesNotApply}%
  \pgfmathtruncatemacro{\tI}{\tH+\nNotPresent}%
  \begin{tikzpicture}[x=1mm,y=1mm]
    \foreach \i in {1,...,72}{%
      \pgfmathsetmacro{\clr}{%
        ifthenelse(\i<=\tA,"colorPresent",%
          ifthenelse(\i<=\tB,"colorPresentBD",%
            ifthenelse(\i<=\tC,"colorPartlyPresent",%
              ifthenelse(\i<=\tD,"colorPartlyPresentBD",%
                ifthenelse(\i<=\tE,"colorLikelyPresent",%
                  ifthenelse(\i<=\tF,"colorLikelyPresentBD",%
                    ifthenelse(\i<=\tG,"colorUnclearFromText",%
                      ifthenelse(\i<=\tH,"colorDoesNotApply", "colorNotPresent")%
                  )%
                )%
              )%
            )%
          )%
        )}%
      \pgfmathtruncatemacro{\col}{int((\i-1)/\rows)}%
      \pgfmathtruncatemacro{\row}{int(mod(\i-1,\rows))}%
      \pgfmathsetmacro{\x}{\col*(\rectW+\hsep)}%
      \pgfmathsetmacro{\y}{-\row*(\rectH+\vsep)}%
      \draw[fill=\clr,draw=\clr,very thin]
        (\x,\y) rectangle ++(\rectW,\rectH);%
    }%
    \pgfmathsetmacro{\gridBottom}{(\rows-1)*(\rectH+\vsep)}%
    \pgfmathsetmacro{\labelX}{\rectW/2}%
    \pgfmathsetmacro{\labelY}{-\gridBottom - 0.7}%
    \draw[colorPresent,thick]
      (\labelX,\labelY) -- ++(0,-2) -- ++(6,0)
      node[black,font=\small,anchor=west]{Present in \pgfmathprintnumber{\nTotalPresent}~papers (\nPresentButDiscussed~discussed)};
  \end{tikzpicture}%
}

\newlength{\presenceLineMargin}
\newcommand{\presenceLine}[9]{%
  \pgfmathtruncatemacro{\nPresent}{#1}%
  \pgfmathtruncatemacro{\nPresentButDiscussed}{#2}%
  \pgfmathtruncatemacro{\nTotalPresent}{#1+#2}%
  \pgfmathtruncatemacro{\nPartlyPresent}{#3}%
  \pgfmathtruncatemacro{\nPartlyPresentButDiscussed}{#4}%
  \pgfmathtruncatemacro{\nLikelyPresent}{#5}%
  \pgfmathtruncatemacro{\nLikelyPresentButDiscussed}{#6}%
  \pgfmathtruncatemacro{\nDoesNotApply}{#7}%
  \pgfmathtruncatemacro{\nUnclearFromText}{#8}%
  \pgfmathtruncatemacro{\nNotPresent}{#9}%

  \pgfmathtruncatemacro{\tA}{\nPresent}%
  \pgfmathtruncatemacro{\tB}{\tA+\nPresentButDiscussed}%
  \pgfmathtruncatemacro{\tC}{\tB+\nPartlyPresent}%
  \pgfmathtruncatemacro{\tD}{\tC+\nPartlyPresentButDiscussed}%
  \pgfmathtruncatemacro{\tE}{\tD+\nLikelyPresent}%
  \pgfmathtruncatemacro{\tF}{\tE+\nLikelyPresentButDiscussed}%
  \pgfmathtruncatemacro{\tG}{\tF+\nUnclearFromText}%
  \pgfmathtruncatemacro{\tH}{\tG+\nDoesNotApply}%
  \pgfmathtruncatemacro{\tI}{\tH+\nNotPresent}%

  \def\rectW{1.55}% mm
  \def\rectH{3.0}% mm
  \def\hsep {0.3}% mm

  \hspace{\presenceLineMargin}%
  \begin{tikzpicture}[x=1mm,y=1mm,baseline=(current bounding box.center)]
    \foreach \i in {1,...,72}{
      \pgfmathsetmacro{\clr}{
        ifthenelse(\i<=\tA,"colorPresent",%
          ifthenelse(\i<=\tB,"colorPresentBD",%
            ifthenelse(\i<=\tC,"colorPartlyPresent",%
              ifthenelse(\i<=\tD,"colorPartlyPresentBD",%
                ifthenelse(\i<=\tE,"colorLikelyPresent",%
                  ifthenelse(\i<=\tF,"colorLikelyPresentBD",%
                    ifthenelse(\i<=\tG,"colorUnclearFromText",%
                      ifthenelse(\i<=\tH,"colorDoesNotApply", "colorNotPresent")%
                    )))))))}
      \pgfmathsetmacro{\x}{(\i-1)*(\rectW+\hsep)}
      \draw[fill=\clr,draw=none] (\x,0) rectangle ++(\rectW,-\rectH);
    }

    \pgfmathsetmacro{\annotX}{0.5*\rectW}
    \pgfmathsetmacro{\annotY}{-0.5*\rectH}
    \draw[colorPresent,thick]
      (\annotX,\annotY) -- ++(0,-3.1) -- ++(5,0)
      node[black,font=\scriptsize,anchor=west]{Present in \pgfmathprintnumber{\nTotalPresent}~papers (\nPresentButDiscussed~discussed)};
  \end{tikzpicture}%
}

\newcommand{\presenceXAxis}[1]{% #1 = number of squares
  \begin{tikzpicture}[x=1mm,y=1mm]
    \def\rectW{1.55}%
    \def\hsep  {0.3}%
    \pgfmathsetmacro{\totalW}{#1*(\rectW+\hsep)-\hsep}

    \draw (0,0) -- (\totalW,0);

    \foreach \k in {0,10,...,100}{
      \pgfmathsetmacro{\x}{\k/100*\totalW}
      \draw (\x,0) -- (\x,-1.5);
      \node[below] at (\x,-1.5) {\k\%};
    }
  \end{tikzpicture}%
}

\newcommand{\donutpie}[3]{ % TODO remove second parameter

  \begin{scope}[shift={#1}]
    \def\radius{1.0}
    \def\inner{0.5}
    \def\start{90}

    \def\colorslist{colorNotPresent, colorDoesNotApply, colorUnclearFromText, colorLikelyPresent, colorPartlyPresent, colorPresent}
    \def\angleslist{#2}

    \newcount\i
    \i=0

    \foreach \color in \colorslist {
      \pgfmathparse{\i+1}
      \xdef\i{\pgfmathresult}
      \pgfmathparse{\angleslist[\i-1]}
      \let\angle=\pgfmathresult

      \filldraw[fill=\color, draw=white] 
        (\start:\inner) arc (\start:\start+\angle:\inner) --
        (\start+\angle:\radius) arc (\start+\angle:\start:\radius) -- cycle;

      \pgfmathparse{\start+\angle} \xdef\start{\pgfmathresult}
    }

    \node (P) at (0,0) {\small #3};
  \end{scope}
}

\newcommand{\getdonutlist}[2]{%
  \def#1{}% leere Liste initialisieren
  \pgfplotstableforeachcolumnelement{Angles}\of\donutdata\as\angle{%
    \ifx#1\empty
      \edef#1{\angle}% erster Eintrag ohne Komma
    \else
      \edef#1{#1,\angle}%
    \fi
  }%
}

\newcommand{\builddonutlist}{%
\getdonutlist{\donutlist}{\donutdata}
}

\newcommand{\pitfallDonut}[2][1]{%
\donutpie{#2}{{\donutlist}}{P#1}%
}

\newcommand{\tikzcircle}[2][red,fill=red]{\tikz[baseline=-0.5ex]\draw[#1,radius=#2] (0,0) circle ;}
\newcommand{\colorcirclemaker}[1]{\raisebox{0.1ex}{\tikzcircle[#1, fill=#1]{2pt}}}

\newcommand{\PIDdatapoisoning}       {1}
\newcommand{\PIDlabelinaccuracy}     {2}
\newcommand{\PIDdataleakage}         {3}
\newcommand{\PIDmodelcollapse}       {4}
\newcommand{\PIDspuriouscorrelations}{5}
\newcommand{\PIDsmallcontext}        {6}
\newcommand{\PIDpromptsensitivity}   {7}
\newcommand{\PIDproxyfallacy}        {8}
\newcommand{\PIDmodelinfo}           {9}

\newcommand{\pitfallname}[1][1]{%
\expandafter\ifstrequal\expandafter{#1}{1}{Data Poisoning via Internet Scraping}{%
\expandafter\ifstrequal\expandafter{#1}{2}{LLM-generated Label Inaccuracy}{%
\expandafter\ifstrequal\expandafter{#1}{3}{Data Leakage}{%
\expandafter\ifstrequal\expandafter{#1}{4}{Model Collapse via Synthetic Training Data}{%
\expandafter\ifstrequal\expandafter{#1}{5}{Spurious Correlations/Unrelated Features}{%
\expandafter\ifstrequal\expandafter{#1}{6}{Context Truncation}{%
\expandafter\ifstrequal\expandafter{#1}{7}{Prompt Sensitivity}{%
\expandafter\ifstrequal\expandafter{#1}{8}{Proxy/Surrogate Fallacy}{%
\expandafter\ifstrequal\expandafter{#1}{9}{Model Ambiguity}{%
'#1' -> nah!%
}}}}}}}}}\xspace%
}

\newcommand{\pitfallnameshort}[1][1]{%
\expandafter\ifstrequal\expandafter{#1}{1}{Data Poisoning}{%
\expandafter\ifstrequal\expandafter{#1}{2}{Label Inaccuracy}{%
\expandafter\ifstrequal\expandafter{#1}{3}{Data Leakage}{%
\expandafter\ifstrequal\expandafter{#1}{4}{Model Collapse}{%
\expandafter\ifstrequal\expandafter{#1}{5}{Spurious Correlations}{%
\expandafter\ifstrequal\expandafter{#1}{6}{Context Truncation}{%
\expandafter\ifstrequal\expandafter{#1}{7}{Prompt Sensitivity}{%
\expandafter\ifstrequal\expandafter{#1}{8}{Surrogate Fallacy}{%
\expandafter\ifstrequal\expandafter{#1}{9}{Model Ambiguity}{%
'#1' -> nah!%
}}}}}}}}}\xspace%
}

\NewDocumentCommand{\Pmodelcollapse}{O{magic}}       {\IfBlankTF{#1}{\pitfallnameshort[\PIDmodelcollapse]}       {\pitfallname[\PIDmodelcollapse]}}
\NewDocumentCommand{\Pdataleakage}{O{magic}}         {\IfBlankTF{#1}{\pitfallnameshort[\PIDdataleakage]}         {\pitfallname[\PIDdataleakage]}}
\NewDocumentCommand{\Psmallcontext}{O{magic}}        {\IfBlankTF{#1}{\pitfallnameshort[\PIDsmallcontext]}        {\pitfallname[\PIDsmallcontext]}}
\NewDocumentCommand{\Pdatapoisoning}{O{magic}}       {\IfBlankTF{#1}{\pitfallnameshort[\PIDdatapoisoning]}       {\pitfallname[\PIDdatapoisoning]}}
\NewDocumentCommand{\Plabelinaccuracy}{O{magic}}     {\IfBlankTF{#1}{\pitfallnameshort[\PIDlabelinaccuracy]}     {\pitfallname[\PIDlabelinaccuracy]}}
\NewDocumentCommand{\Pmodelinfo}{O{magic}}           {\IfBlankTF{#1}{\pitfallnameshort[\PIDmodelinfo]}           {\pitfallname[\PIDmodelinfo]}}
\NewDocumentCommand{\Ppromptsensitivity}{O{magic}}   {\IfBlankTF{#1}{\pitfallnameshort[\PIDpromptsensitivity]}   {\pitfallname[\PIDpromptsensitivity]}}
\NewDocumentCommand{\Pproxyfallacy}{O{magic}}        {\IfBlankTF{#1}{\pitfallnameshort[\PIDproxyfallacy]}        {\pitfallname[\PIDproxyfallacy]}}
\NewDocumentCommand{\Pspuriouscorrelations}{O{magic}}{\IfBlankTF{#1}{\pitfallnameshort[\PIDspuriouscorrelations]}{\pitfallname[\PIDspuriouscorrelations]}}

\newcommand{\pitfallPresenceGrid}[1][1]{%
\expandafter\ifstrequal\expandafter{#1}{1}{\presenceGrid{3}{0}{1}{0}{17}{0}{30}{1}{20}}{%
\expandafter\ifstrequal\expandafter{#1}{2}{\presenceGrid{3}{5}{2}{4}{1}{0}{6}{0}{51}}{%
\expandafter\ifstrequal\expandafter{#1}{3}{\presenceGrid{8}{7}{3}{2}{26}{1}{5}{5}{15}}{%
\expandafter\ifstrequal\expandafter{#1}{4}{\presenceGrid{10}{0}{2}{0}{1}{0}{27}{1}{31}}{%
\expandafter\ifstrequal\expandafter{#1}{5}{\presenceGrid{3}{1}{0}{1}{17}{0}{6}{8}{36}}{%
\expandafter\ifstrequal\expandafter{#1}{6}{\presenceGrid{8}{6}{0}{4}{6}{0}{3}{11}{34}}{%
\expandafter\ifstrequal\expandafter{#1}{7}{\presenceGrid{10}{3}{3}{2}{5}{0}{13}{1}{35}}{%
\expandafter\ifstrequal\expandafter{#1}{8}{\presenceGrid{11}{3}{17}{2}{1}{0}{0}{1}{37}}{%
\expandafter\ifstrequal\expandafter{#1}{9}{\presenceGrid{53}{0}{9}{0}{0}{0}{1}{0}{9}}{%
'#1' -> nah!%
}}}}}}}}}%
}

\iffalse % new or old
\iftrue % left or right
\newenvironment{pitfall}[1][1]{%
\noindent%
\newcommand{\pid}{#1}%
\begin{minipage}[t]{0.3\columnwidth}%
    \begin{flushleft}
    \begin{tikzpicture}[baseline={([yshift=3.8em]P.base)}]%
        \pgfplotstableread[col sep=comma]{results/P\pid.csv}\donutdata%
        \builddonutlist%
        \donutpie{(0,0)}{{\donutlist}}{P\pid}%
    \end{tikzpicture}%
    \end{flushleft}~\\[-10pt]
\end{minipage}%
\begin{minipage}[t]{0.7\columnwidth}%
    \vspace*{1mm}\boldpar{\pitfallname[#1]}%
}{%
~\\
\end{minipage}~%
}
\else
\newenvironment{pitfall}[1][1]{%
\noindent%
\newcommand{\pid}{#1}%
\begin{minipage}[t]{0.7\columnwidth}%
    \vspace*{1mm}\boldpar{P\pid---\pitfallname[#1]}%
}{%
\end{minipage}~%
\begin{minipage}[t]{0.3\columnwidth}%
    \begin{flushright}
    \begin{tikzpicture}[baseline={([yshift=3.8em]P.base)}]%
        \pgfplotstableread[col sep=comma]{results/P\pid.csv}\donutdata%
        \builddonutlist%
        \donutpie{(0,0)}{{\donutlist}}{P\pid}%
    \end{tikzpicture}%
    \end{flushright}
\end{minipage}%
}
\fi

\else

\NewDocumentEnvironment{pitfall}{O{1} O{showstats}}{%
\noindent%
\newcommand{\pid}{#1}%
    \IfBlankTF{#2}{%
    ~\\%
    }{%
    \begin{minipage}[c]{\columnwidth}%
    \vspace*{2mm}
    }%
    \mbox{\boldpar{P\pid---\expandafter\pitfallname\expandafter[#1]}}%
}{%
    \IfBlankTF{#2}{}{%
    \begin{center}%
	\expandafter\pitfallPresenceGrid\expandafter[\pid]\vspace*{1mm}%
    \end{center}%
    \end{minipage}%
    }%
}

\fi

\newcommand{\mc}[1]{\multicolumn{1}{c}{#1}}
\newcommand{\mcb}[1]{\multicolumn{1}{c}{\bfseries #1}}
\newcommand{\mr}[2][1]{\multirow{#1}{*}{\makecell{#2}}}

\begin{document}

\title{\centering Chasing Shadows:\\Pitfalls in LLM Security Research}

\author{
\IEEEauthorblockN{
Jonathan Evertz$^{*\,\dagger}$,
Niklas Risse$^{*\,\ddagger}$,
Nicolai Neuer$^{\S}$,
Andreas Müller$^{\P}$,
Philipp Normann$^{\|}$, \\
Gaetano Sapia$^{\ddagger}$, 
Srishti Gupta$^{\#}$,
David Pape$^{\dagger}$,
Soumya Shaw$^{\dagger}$,
Devansh Srivastav$^{\dagger}$,  \\
Christian Wressnegger$^{\S}$,
Erwin Quiring$^{\diamond}$,
Thorsten Eisenhofer$^{\dagger}$,
Daniel Arp$^{\|}$,
Lea Schönherr$^{\dagger}$
\\[0.6em]
}
\IEEEauthorblockA{
$^{\dagger}$ CISPA Helmholtz Center for Information Security \quad
$^{\ddagger}$ Max Planck Institute for Security and Privacy \\[0.3em]
$^{\S}$ Karlsruhe Institute of Technology \quad
$^{\P}$ Ruhr University Bochum \quad
$^{\|}$ TU Wien \\[0.3em]
$^{\#}$ Sapienza University of Rome \quad
$^{\diamond}$ \_fbeta\\[0.6em]
$^{*}$ Equal contribution
}
}

\IEEEoverridecommandlockouts
\makeatletter\def\@IEEEpubidpullup{6.5\baselineskip}\makeatother
\IEEEpubid{\parbox{\columnwidth}{
		Network and Distributed System Security (NDSS) Symposium 2026\\
		23-27 February 2026, San Diego, CA, USA\\
		ISBN 979-8-9919276-8-0\\
		https://dx.doi.org/10.14722/ndss.2026.241749\\
		www.ndss-symposium.org
}
\hspace{\columnsep}\makebox[\columnwidth]{}}

\maketitle

\pagestyle{plain}

\begin{abstract}
Large language models (LLMs) are increasingly prevalent in security research. Their unique characteristics, however, introduce challenges that undermine established paradigms of reproducibility, rigor, and evaluation. Prior work has identified common pitfalls in traditional machine learning research, but these studies predate the advent of LLMs. In this paper, we identify \emph{nine} common pitfalls that have become (more) relevant with the emergence of LLMs and that can compromise the validity of research involving them. These pitfalls span the entire computation process, from data collection, pre-training, and fine-tuning to prompting and evaluation.

We assess the prevalence of these pitfalls across all 72 peer-reviewed papers published at leading Security and Software Engineering venues between 2023 and 2024. We find that every paper contains at least one pitfall, and each pitfall appears in multiple papers. Yet only 15.7\% of the present pitfalls were explicitly discussed, suggesting that the majority remain unrecognized. To understand their practical impact, we conduct four empirical case studies showing how individual pitfalls can mislead evaluation, inflate performance, or impair reproducibility. Based on our findings, we offer actionable guidelines to support the community in future work.
\end{abstract}

% \begin{IEEEkeywords}
% Large language models, Machine learning, Pitfalls, Security
% \end{IEEEkeywords}

\section{Introduction}
\label{sec:introduction}

The intersection of large language models (LLMs) and security has become a fast-growing and influential line of research. Their capacity for easy adaptation and deployment makes LLMs powerful instruments for security-critical applications, including vulnerability detection~\cite{vulndet5, vulndet2, vulndet4, fuzzing8}, code analysis~\cite{fang_code, zhou_vulndet, garak, purba_vulndet}, and automated code repair~\cite{vulnrep1, vulnrep2, vulnrep3, vulnrep4}. At the same time, it is crucial to understand their inherent capabilities for resisting attacks: models struggle to distinguish commands from data~\cite{zverev2025can}, to identify context-dependent sensitive content~\cite{evertz2024whispers, hankache2025evaluatingsensitivityllmsprior}, and are generally vulnerable to prompt injection~\cite{liu_pi} and jailbreak attacks~\cite{yu_jailbreaks}. For LLMs to be trusted in critical applications, both in research and in practice, their implementations must be rigorous, reproducible, and methodologically sound.

Prior work has identified common pitfalls in traditional machine learning research~\cite{arp_dos_2022, carlini2019evaluating}, offering valuable guidance for designing sound experiments. However, these studies predate the emergence of modern LLMs.
Language models have fundamentally reshaped the stages of the machine learning workflow and introduced new risks: data collection (\eg, poisoning via large-scale web scraping), pre-training (\eg, inadvertent data leakage), fine-tuning (\eg, reliance on synthetic data), prompt engineering (\eg, context sensitivity), and evaluation (\eg, unstable API behavior). As a result, previous studies~\cite{arp_dos_2022, carlini2019evaluating} do not fully capture the complexity and fragility of current LLM practices.

In this paper, we bridge this gap by identifying nine pitfalls that commonly occur in LLM security research. These pitfalls span the entire development pipeline of LLMs. They reflect both structural changes in how models are built and used, as well as the unique challenges introduced by the scale, opacity, and natural language interface of LLMs. Some pitfalls are entirely new, such as \emph{model collapse} caused by training on synthetic data or unpredictable model behavior due to \emph{prompt sensitivity}. Others represent familiar concerns, such as \emph{data leakage} or \emph{spurious correlations}, but their implications change and intensify in the context of LLMs.

To assess how widespread these pitfalls are, we conducted a prevalence study across all LLM-centric research papers (72 in total) published between January 2023 and December 2024 at leading conferences in Security (IEEE S\&P, NDSS, ACM CCS, and USENIX Security Symposium) and Software Engineering (IEEE/ACM ICSE, ACM ISSTA, ACM FSE, and IEEE/ACM ASE). With a team of 15 researchers, we collaboratively developed detailed labeling guidelines and reviewed each paper independently using a systematic and uniform agreement process.

\begin{figure*}[!t]
    \centering
    % vordefinierte Maße
\tikzset{
  block/.style={
    rectangle, draw=black, thick, minimum width=3.0cm, minimum height=1.2cm, align=center, fill=white
  },
  arrow/.style={
    ->, thick, >=stealth
  },
  groupbg/.style={
    draw=black,
    thick,
    dashed,
    fill=gray!20,
    rounded corners,
    inner sep=0.5cm
  },
  stage/.style = {font=\small\bfseries, above=0.2mm of #1},
  node distance=2cm
}

    \begin{tikzpicture}
    
      \newcommand{\basepath}{results/pipeline/}
    
      % Rechtecke
      \node[block] (block1) {Data Collection\\and Labeling};
      \node[block, right=of block1] (block2) {Pre-Training};
      \node[block, right=of block2] (block3) {Fine-tuning \\ and Alignment};
      \node[block, right=of block3] (block4) {Prompt Engineering};
      \node[block, right=of block4] (block5) {Evaluation};

      % Stage labels
      \node[stage=block1] {Stage 1};
      \node[stage=block2] {Stage 2};
      \node[stage=block3] {Stage 3};
      \node[stage=block4] {Stage 4};
      \node[stage=block5] {Stage 5};

      % Title
      %\node[font=\bfseries] at ($($(block1.north)!0.5!(block5.north)$)+(0,1.2)$) {Large Language Model Pipeline};
      \node[font=\bfseries] at ($(block1.north)+(0.5,1.2)$) {Large Language Model Pipeline};

      % Hintergrundrechteck um alle Blöcke
      \begin{pgfonlayer}{background}
        \node[groupbg, fit=(block1)(block5), name=legendfit, yshift=2mm] {};
      \end{pgfonlayer}

      % Pfeile mit "arrows"-Bibliothek
      \draw[arrow] (block1) -- (block2);
      \draw[arrow] (block2) -- (block3);
      \draw[arrow] (block3) -- (block4);
      \draw[arrow] (block4) -- (block5);

      % Senkrechte Linien nach unten ab 1/5 Breite
      \foreach \blk in {block1, block2, block3, block4, block5} {
        \draw[black, thick]
          ($(\blk.south west)+(0.6,0)$) -- ++(0,-0.7);
      }

      \pgfplotstableread[col sep=comma]{\basepath P4.csv}\donutdata
      \builddonutlist % Hier wird die Liste gebaut
      \pitfallDonut[1]{($(block1.south west)+(0.6,-1.7)$)}

      \pgfplotstableread[col sep=comma]{\basepath P1.csv}\donutdata
      \builddonutlist
      \pitfallDonut[4]{($(block3.south west)+(0.6,-1.7)$)}

      \pgfplotstableread[col sep=comma]{\basepath P2.csv}\donutdata
      \builddonutlist
      \pitfallDonut[3]{($(block2.south west)+(0.6,-1.7)$)}

      \pgfplotstableread[col sep=comma]{\basepath P3.csv}\donutdata
      \builddonutlist
      \pitfallDonut[6]{($(block4.south west)+(0.6,-1.7)$)}

      \pgfplotstableread[col sep=comma]{\basepath P8.csv}\donutdata
      \builddonutlist
      \pitfallDonut[8]{($(block5.south west)+(0.6,-1.7)$)}

      % Zweite senkrechte Linien unter dem ersten Donut
      \foreach \blk in {block1, block3, block4, block5} {
        \draw[black, thick]
            ($(\blk.south west)+(0.6,-2.7)$) -- ++(0,-0.7);
      }

      % Zweite Reihe Donuts:

      \pgfplotstableread[col sep=comma]{\basepath P5.csv}\donutdata
      \builddonutlist
      \pitfallDonut[2]{($(block1.south west)+(0.6,-4.4)$)}

      \pgfplotstableread[col sep=comma]{\basepath P9.csv}\donutdata
      \builddonutlist
      \pitfallDonut[5]{($(block3.south west)+(0.6,-4.4)$)}

      \pgfplotstableread[col sep=comma]{\basepath P7.csv}\donutdata
      \builddonutlist
      \pitfallDonut[7]{($(block4.south west)+(0.6,-4.4)$)}

      \pgfplotstableread[col sep=comma]{\basepath P6.csv}\donutdata
      \builddonutlist
      \pitfallDonut[9]{($(block5.south west)+(0.6,-4.4)$)}

      % Beschriftungen setzen
      \node[anchor=west] at ($(block1.south west)+(0.6,-1.7)+(1.1,0)$) {\Pdatapoisoning[]};
      \node[anchor=west] at ($(block1.south west)+(0.6,-4.4)+(1.1,0)$) {\Plabelinaccuracy[]};

      \node[anchor=west] at ($(block3.south west)+(0.6,-1.7)+(1.1,0)$) {\Pmodelcollapse[]};
      \node[anchor=west, align=left] at ($(block2.south west)+(0.6,-1.7)+(1.1,0)$) {\Pdataleakage[]};

      \node[anchor=west, align=left, text width=20mm] at ($(block3.south west)+(0.6,-4.4)+(1.1,0)$) {\Pspuriouscorrelations[]};

      \node[anchor=west, align=left, text width=20mm] at ($(block4.south west)+(0.6,-1.7)+(1.1,0)$) {\Psmallcontext[]};
      \node[anchor=west, align=left, text width=20mm] at ($(block4.south west)+(0.6,-4.4)+(1.1,0)$) {\Ppromptsensitivity[]};

      \node[anchor=west, align=left, text width=20mm] at ($(block5.south west)+(0.6,-1.7)+(1.1,0)$) {\Pproxyfallacy[]};
      \node[anchor=west, align=left, text width=20mm] at ($(block5.south west)+(0.6,-4.4)+(1.1,0)$) {\Pmodelinfo[]};

        \newif\ifusebox
        \useboxfalse

        \ifusebox
            % Legende: Kategorien mit discussed
            \begin{scope}[yshift=-8.5cm] 
              % Fixe Y-Koordinate für die Linie
              \coordinate (legendY) at (0,0.6);
            
              % Verwende die X-Koordinaten von legendfit und die Y-Koordinate von legendY
              \path (legendfit.north west |- legendY) coordinate (legendLineStart);
              \path (legendfit.north east |- legendY) coordinate (legendLineEnd);
              \draw[thick] (legendLineStart) -- (legendLineEnd);
              
              \begin{scope}[shift={(legendLineStart)}, yshift=-0.8cm]
                  % Spalte 1
                  \node at (0,0) [fill=colorNotPresent, draw=black, minimum width=0.4cm, minimum height=0.4cm] {};
                  \node[anchor=west] at (0.6,0) {Not Present};
            
                  \node at (0,-1) [fill=colorPartlyPresentBD, draw=black, minimum width=0.4cm, minimum height=0.4cm] {};
                  \node[anchor=west] at (0.6,-1) {Partly Present (Discussed)};
            
                  % Spalte 2
                  \node at (5.5,0) [fill=colorDoesNotApply, draw=black, minimum width=0.4cm, minimum height=0.4cm] {};
                  \node[anchor=west] at (6.1,0) {Does Not Apply};
            
                  \node at (5.5,-1) [fill=colorPartlyPresent, draw=black, minimum width=0.4cm, minimum height=0.4cm] {};
                  \node[anchor=west] at (6.1,-1) {Partly Present};
            
                  % Spalte 3
                  \node at (11,0) [fill=colorUnclearFromText, draw=black, minimum width=0.4cm, minimum height=0.4cm] {};
                  \node[anchor=west] at (11.6,0) {Unclear From Text};
            
                  \node at (11,-1) [fill=colorPresentBD, draw=black, minimum width=0.4cm, minimum height=0.4cm] {};
                  \node[anchor=west] at (11.6,-1) {Present (Discussed)};
            
                  % Spalte 4
                  \node at (16.5,0) [fill=colorLikelyPresentBD, draw=black, minimum width=0.4cm, minimum height=0.4cm] {};
                  \node[anchor=west] at (17.1,0) {Likely Present (Discussed)};
            
                  \node at (16.5,-1) [fill=colorPresent, draw=black, minimum width=0.4cm, minimum height=0.4cm] {};
                  \node[anchor=west] at (17.1,-1) {Present};
            
                  % Spalte 5 (nur oben)
                  \node at (22,0) [fill=colorLikelyPresent, draw=black, minimum width=0.4cm, minimum height=0.4cm] {};
                  \node[anchor=west] at (22.6,0) {Likely Present};
                    %\end{tikzpicture}
                  %};
              \end{scope}
            \end{scope}
        \else
            % Legende: Kategorien ohne discussed
            \begin{scope}[yshift=-7cm] 

              % Fixe Y-Koordinate für die Linie
              \coordinate (legendY) at (0,0.6);
            
              % Verwende die X-Koordinaten von legendfit und die Y-Koordinate von legendY
              \path (legendfit.north west |- legendY) coordinate (legendLineStart);
              \path (legendfit.north east |- legendY) coordinate (legendLineEnd);
              \draw[thick] (legendLineStart) -- (legendLineEnd);
              
              \begin{scope}[shift={(legendLineStart)}, xshift=0.85cm, yshift=-0.5cm]
                  % Spalte 1
                  \node at (0,0) [fill=colorNotPresent, draw=black, minimum width=0.4cm, minimum height=0.4cm] {};
                  \node[anchor=west] at (0.5,0) {Not Present};
                 
                  % Spalte 2
                  \node at (4.2,0) [fill=colorDoesNotApply, draw=black, minimum width=0.4cm, minimum height=0.4cm] {};
                  \node[anchor=west] at (4.5,0) {Does Not Apply};
                    
                  % Spalte 3
                  \node at (8.4,0) [fill=colorUnclearFromText, draw=black, minimum width=0.4cm, minimum height=0.4cm] {};
                  \node[anchor=west] at (8.7,0) {Unclear From Text};

                  % Spalte 4
                  \node at (12.6,0) [fill=colorLikelyPresent, draw=black, minimum width=0.4cm, minimum height=0.4cm] {};
                  \node[anchor=west] at (12.9,0) {Likely Present};

                  % Spalte 4
                  \node at (16.8,0) [fill=colorPartlyPresent, draw=black, minimum width=0.4cm, minimum height=0.4cm] {};
                  \node[anchor=west] at (17.1,0) {Partly Present};

                  % Spalte 5
                  \node at (21,0) [fill=colorPresent, draw=black, minimum width=0.4cm, minimum height=0.4cm] {};
                  \node[anchor=west] at (21.3,0) {Present};
                    %\end{tikzpicture}
                  %};
              \end{scope}
            \end{scope}
        \fi
    
    \end{tikzpicture}
\end{document}
    \caption{%
    Typical LLM pipeline, divided into its key stages. Each stage can introduce LLM-specific pitfalls that may distort evaluation, inflate reported performance, or undermine reproducibility. Colors indicate the prevalence of each pitfall according to our prevalence study (\S\ref{sec:prevalence_assessment}).
    } 
    \label{fig:llm_pipeline}
\end{figure*}
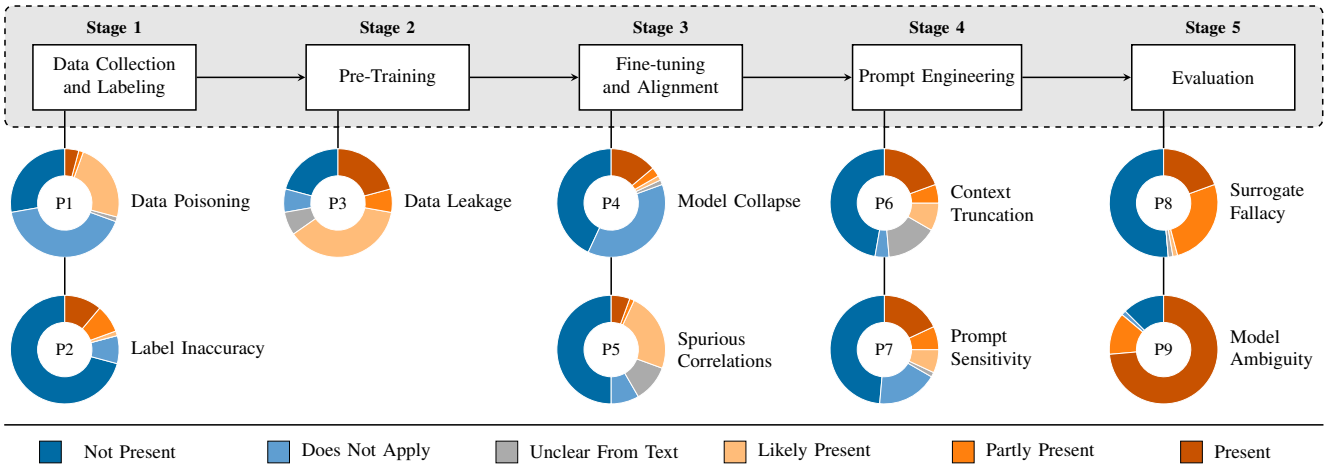

To our surprise, \emph{every} paper in our study contains at least one of the nine identified pitfalls as part of its main contribution. Five pitfalls, namely \Pdataleakage[]\ (P\PIDdataleakage), \Psmallcontext[]\ (P\PIDsmallcontext), \Ppromptsensitivity[]\ (P\PIDpromptsensitivity), \Pproxyfallacy[]\ (P\PIDproxyfallacy), and \Pmodelinfo[]\ (P\PIDmodelinfo), appear in more than 20\% of the papers. Prevalence also varies across research topics: papers on \emph{Vulnerability Repair and Detection} show the highest average rate of pitfalls per paper (23--28\%), while studies on \emph{Fuzzing} and \emph{Secure Code Generation} contain fewer pitfalls on average (15--18\%).

Beyond identifying prevalence, we conduct four case studies that experimentally demonstrate how the identified pitfalls can mislead evaluation, inflate performance, or hinder reproducibility. Concretely, we examine the impact of (i) \Pmodelinfo[]---snapshot versions and quantization affect robustness against jailbreaks and significantly alter precision/recall, thereby affecting reproducibility; (ii) \Pdataleakage[]---leaking 20\% of test data into the fine-tuning process raises the F1 score by $\approx$~0.08--0.11 with increases that grow almost linearly as leakage increases; (iii) \Psmallcontext[]---about 49\% of vulnerable functions exceed common context windows of 512 tokens (29\% $>1024$ tokens), removing essential information and distorting evaluation; (iv) \Pmodelcollapse[]---recursive self-training in code generation increases perplexity across generations, leading to degradation and instability.

Based on our results, we provide concrete guidelines and recommendations for each identified pitfall. \emph{Importantly, our goal is not to assign blame.} Pitfalls can occur in carefully conducted research, often without being explicitly recognized. Our aim is to support the community by raising awareness and offering actionable recommendations.

\boldpar{Contributions} We make the following key contributions:
\begin{itemize}
    \item \textbf{Identification of LLM pitfalls.} We identify nine pitfalls that frequently arise in LLM-based security research spanning the entire computation pipeline from data collection to evaluation. We describe each pitfall, explain why it occurs, and discuss its potential impact on validity and reproducibility.
    \item \textbf{Prevalence study.} We assess the prevalence of all nine pitfalls across 72 peer-reviewed papers published between January 2023 and December 2024 at leading Security and Software Engineering venues. We find that every paper suffers from at least one pitfall.
    \item \textbf{Impact analysis.} We conduct four case studies that empirically demonstrate how individual pitfalls can distort evaluation, inflate performance metrics, or compromise reproducibility.
    \item \textbf{Guidelines and recommendations.} We present guidelines and recommendations for each identified pitfall and maintain a living appendix containing up-to-date information on best practices for preventing pitfalls, accessible at \url{https://llmpitfalls.org}.
\end{itemize}

We provide all code, datasets, and complete reproduction instructions at \url{https://github.com/dormant-neurons/llm-pitfalls}. %A living appendix with up-to-date information and guidelines for preventing pitfalls is available at \url{https://llmpitfalls.org}.
\section{Pitfalls Across the LLM Pipeline} \label{sec:methodology}

%\subsection{Traditional ML pitfalls} 
Research on pitfalls in machine learning is not new. Prior work has systematically analyzed common failure modes in traditional pipelines~\cite{arp_dos_2022, carlini2019evaluating}. While related in spirit, the pitfalls we investigate arise from the specific characteristics of LLM research. %The scale of training data, limited transparency of proprietary models, and natural language as both input and interface introduce pitfalls that are not well captured in earlier taxonomies. 
Our aim is not to replicate existing analyses, but to complement them with a targeted examination of challenges that are unique to or substantially intensified in LLMs.

LLMs have significantly changed the way machine learning systems are trained, adapted, and evaluated. Their development pipelines involve multiple interdependent stages: large-scale data collection and labeling, pre-training, fine-tuning, prompt design, and evaluation.
In the following sections, we outline the structure of this pipeline and describe the potential practical risks that can arise at each stage. This discussion forms the foundation for the nine pitfalls we identify. Figure~\ref{fig:llm_pipeline} illustrates the overall pipeline and the corresponding pitfalls.

\medskip

\subsection{Stage 1: Data Collection and Labeling} 
LLMs require large-scale datasets to train effectively, often relying on content scraped from the Internet, which is far more extensive and noisy than the more curated datasets used in classical machine learning.

\begin{pitfall}[\PIDdatapoisoning][]  
The sheer scale of this data makes curation increasingly difficult, creating favorable conditions for \emph{\Pdatapoisoning[]}.
Without sufficient precautions, data scraped from the Internet opens the door to data poisoning attacks, where malicious or biased content can be subtly inserted into the training data without detection. Although not new to machine learning, the scale of LLMs exacerbates this risk. Platforms like GitHub or Reddit, where anyone can anonymously upload data, make it especially difficult to ensure quality and safety at scale.
\end{pitfall}

\begin{pitfall}[\PIDlabelinaccuracy][]
At the same time, the demand for labeled data during fine-tuning and evaluation has led to the widespread use of models themselves to generate labels, a practice known as LLM-as-a-judge~\cite{zheng_judging_2023}.
Although this approach can address the problem of generating labels in scenarios in which human-curated labeling is time-intensive or costly, it also introduces the risk of \mbox{\emph{\Plabelinaccuracy[]}}.

The reliability of experimental results depends heavily on the quality of data and labeling. Since learning-based methods rely on accurate labels, any label errors or instabilities can degrade performance. This is particularly relevant when LLMs are used for automated annotation, where outputs may appear correct yet still contain subtle flaws.
\end{pitfall}

\medskip

\subsection{Stage 2: Pre-Training}

LLMs are typically pre-trained on large-scale datasets to capture general language patterns. Because these corpora are often assembled from broad internet scrapes with limited transparency, they increase the risk of data leakage.

\begin{pitfall}[\PIDdataleakage][]
Many model providers no longer disclose detailed information about the data used for training. For instance, \model{\GPTfour} was trained on ``a variety of licensed, created, and publicly available data sources''~\cite{openai_gpt4_system_card_2023,openai_gpt4_tech_2023}, without further specification. This lack of transparency increases the risk of \emph{\Pdataleakage[]}, where evaluation data may inadvertently overlap with training inputs.

 While creating traditional machine learning models with disjoint data splits is straightforward, the nature of LLMs complicates this. As LLMs require massive datasets, fine-tuning foundational models is a widespread practice. Foundational models may have been trained on (a subset of) samples from the test set. This risk is particularly high in LLMs, where many datasets overlap with pre-training sources like GitHub, Wikipedia, and Reddit~\cite{golem_reddit_anthropic, golem_reddit_ki}.
\end{pitfall}

\medskip

\subsection{Stage 3: Fine-tuning and Alignment} 

During fine-tuning or alignment, LLMs are adapted for specific downstream tasks and user-facing behaviors. In this stage, models become particularly susceptible to issues where evaluation data may inadvertently overlap with pre-training data, as well as the reliance on synthetic or LLM-generated~data.

\begin{pitfall}[\PIDmodelcollapse][] 
Fine-tuning and alignment frequently rely on data that was itself generated by LLMs, introducing the risk of \emph{\Pmodelcollapse[]}~\cite{shumailov_ai_2024}. 

LLMs require vast amounts of data, and this need has increased substantially. To address data scarcity, synthetic data generated by other LLMs is frequently used. However, training on such data can lead to less diverse outputs, higher error rates, and model collapse over generations. While countermeasures exist, they are difficult to implement in practice~\cite{pmlr-v235-gillman24a, kossale_mode_2022, alemohammad_self-consuming_2023}.
\end{pitfall}

\begin{pitfall}[\PIDspuriouscorrelations][] 
Additionally, \emph{\Pspuriouscorrelations[]}, already known as an issue in traditional ML, becomes especially problematic for LLMs because their massive parameter space allows them to memorize and exploit subtle, non-causal patterns. 

Spurious correlations are artifacts that correlate with task labels but are not actually related to the underlying task. Such features may arise from selection bias in the training data or from shortcuts the model learns using unrelated patterns. This can lead to misleading performance and poor generalization.
\end{pitfall}

\medskip
\subsection{Stage 4: Prompt Engineering} 
Prompting has emerged as a new form of control in LLMs. It effectively acts as a tunable hyperparameter that controls the model's behavior, and its design can influence the model’s outputs and overall performance.

\begin{pitfall}[\PIDsmallcontext][] 
%
%With the shift to autoregressive architectures, models now operate within fixed-size context windows, introducing the issue of \emph{\Psmallcontext}.
%\medskip
An LLM's context size refers to the maximum number of tokens it can process at once. Because LLMs are stateless, all relevant information must be included in the current context. If inputs exceed this limit, they are truncated, potentially omitting critical information and reducing performance.
\end{pitfall}

\begin{pitfall}[\PIDpromptsensitivity][] 
Prompting also introduces the issue of \emph{\Ppromptsensitivity[]}, where minor changes in phrasing can lead to drastically different outputs and where different models can have distinct prompt preferences.

Language models are often fine-tuned to follow instruction-based inputs with specific formatting styles. While this allows the language model to adapt to desired tasks during runtime, the performance and functionality of the model also rely on the quality of the instructions and the correct input format. This results in differences across evaluations if inputs are formatted for a specific LLM but evaluated on other models, or are not expressive enough for the task~\cite{cao2024on, pryzant2023automatic}.
\end{pitfall}

\medskip
\subsection{Stage 5: Evaluation} 
As LLMs are often accessed through APIs and web interfaces, with several distinct model versions that differ in architecture, parameter count, or quantization. This ambiguity creates risks in reproducing results, as identifying the precise model instance becomes difficult or even infeasible.

\begin{pitfall}[\PIDproxyfallacy][] 
A single model name, such as \model{ChatGPT}, may refer to multiple underlying snapshots with different architectures, sizes, or quantization levels. This creates the risk of the \emph{\Pproxyfallacy},  where conclusions are drawn from models that are not representative of those actually~used.

New LLMs are released frequently, driving the race for the ``most state-of-the-art’’ model. These models vary in size and task specialization. Although models of the same class often share architecture and training data, differences in size or context window can significantly affect performance and behavior. Discrepancies are even greater across model classes, for example, between the open-source \model{Llama} models~\cite{touvron_llama_2023, llama3_website, llama3herd} and the commercial \model{\GPTmodel} family~\cite{openai_chatgpt}. Without experimental validation, it is generally not possible to make claims between different model sizes and families.
\end{pitfall}

\begin{pitfall}[\PIDmodelinfo][] 
The ambiguity of different architectures, sizes, or quantization levels also leads to \emph{\Pmodelinfo[]} risks, where determining the exact model instance used to reproduce a result is difficult or even impossible.

Both open-source models on platforms like \entity{Huggingface}\cite{huggingface} and proprietary models like \model{\GPTapp}\cite{openai_chatgpt}, \model{Claude}\cite{anthropic_claude}, or \model{Gemini}\cite{google_gemini} are regularly updated. Minor updates are often noted only in changelogs and may not result in a new major version. A single model specifier can therefore refer to different internal versions. Because even small changes in tokenizers or system prompts can affect model behavior, specifying the exact snapshot (\eg, endpoint version, commit ID, or access date) is essential. For open-source models, the quantization level is often an additional source of variation and should be specified (cf., \S\ref{sec:no_detailed_model_version}).
\end{pitfall}
\section{Prevalence of Pitfalls} \label{sec:prevalence_assessment}

Having identified nine pitfalls specific to the LLM computation pipeline, we now assess how frequently these issues appear in current LLM security research.
We begin by outlining the paper selection process that underpins our study (\S\ref{sec:paper-collection}) as well as the methodology used to evaluate each paper (\S\ref{sec:review_methodology}). We then present the results per pitfall (\S\ref{sec:prevalence_assessment:prevalence_results}), followed by a broader discussion (\S\ref{sec:prevalence_assessment:analysis}).

\subsection{Paper Collection}
\label{sec:paper-collection}

As the basis for our study, we consider papers published between January 2023 and December 2024 at leading conferences in Security and Software Engineering. 

\begin{itemize}
    \item For Security, we consider the IEEE Symposium on Security and Privacy (IEEE S\&P), the Network and Distributed System Security Symposium (NDSS), the ACM Conference on Computer and Communications Security (ACM CCS), and the USENIX Security Symposium. 
    \item For Software Engineering, we consider the IEEE/ACM International Conference on Software Engineering (IEEE/ACM ICSE), the ACM International Conference on the Foundations of Software Engineering (ACM FSE), the ACM SIGSOFT International Symposium on Software Testing and Analysis (ACM ISSTA), and the IEEE/ACM International Conference on Automated Software Engineering (IEEE/ACM ASE). 
\end{itemize}

For all conferences, we scraped the full set of papers and applied a two-stage filtering process to identify relevant studies, as detailed next.

\boldpar{Selection Criteria} To identify relevant papers, we establish two primary selection criteria centered on the use of LLMs in security-relevant contexts.
Specifically, papers were considered in scope if they (1) evaluate the security or safety of LLMs, or (2) apply LLMs to security-related tasks such as vulnerability detection or secure code generation. Importantly, the use of the LLM had to be integral to a paper's methodology, meaning the core contribution would not have been possible in its current form without them.

\begin{figure}[t]
\centering
\resizebox{\linewidth}{!}{
\begin{tikzpicture}
\pie[
    text=pin,
    sum=auto,
    after number=\%,
    draw=white,
    style={ultra thick},
    color={
        colorLLMForSecurity, colorLLMForSecurity, colorLLMForSecurity, colorLLMForSecurity,  % LLMs for security
        colorSecurityInLLM, colorSecurityInLLM, % Security/safety of LLMs
    }
]{
    22.2/Vulnerability Detection (15),
    8.3/Vulnerability Repair (6),
    11.2/Fuzzing (8),
    5.6/Secure Code Generation (5),
    33.3/Security in LLM (24),
    19.4/GenAI Safety (14)
}
\end{tikzpicture}
}
\caption{Distribution of the 72 selected papers across topics. The top half, light blue segments (\colorcirclemaker{colorLLMForSecurity}) represent research that uses LLMs to address security problems. The bottom half, dark blue segments (\colorcirclemaker{colorSecurityInLLM}) correspond to research focused on the security and safety of LLMs themselves.}
\label{fig:paper-categories}
\end{figure}
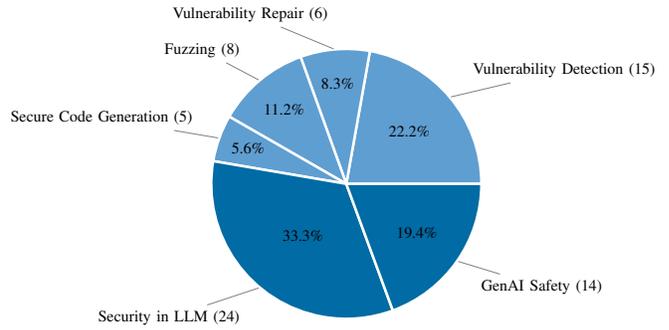

To guide the search, we define a list of LLM-relevant keywords: \emph{LLM}, \emph{LLMs}, \emph{large language model}, \emph{language model}, \emph{\GPTmodel}, \emph{\GPTapp}, \emph{transformer}, \emph{pre-trained}, \emph{foundation model}, \emph{prompt}, \emph{learning}. 
We applied these keywords to search the titles and abstracts of all papers published in the selected venues using multiple specialized tools for literature search: IEEE Xplore Advanced Search\footnote{\href{https://ieeexplore.ieee.org/search/advanced}{IEEE Xplore Advanced Search}}, ACM DL Advanced Search\footnote{\href{https://dl.acm.org/search/advanced}{ACM DL Advanced Search}}, and Google Scholar Advanced Search\footnote{\href{https://scholar.google.com/}{Google Scholar Advanced Search}}.

This process yielded over 1.000 candidate papers, which we manually reviewed for relevance. As a first step, we excluded papers that clearly did not fit our scope. For example, we removed the software engineering paper \emph{Large Language Models are Few-Shot Summarizers: Multi-Intent Comment Generation via In-Context Learning}~\cite{2025_mingyang_few_shot_summarizers}, which focuses on code summarization without any direct security aspect.
Second, all remaining abstracts, and when necessary, the full texts, were manually reviewed to assess whether each paper met the inclusion criteria.

\boldpar{Selected Papers} Our collection process resulted in a final set of 72 papers. These papers span a wide range of topics within the field of language models and security: 
\emph{Security in LLM}~\cite{secinllm1, secinllm2, secinllm3, secinllm4, secinllm5, secinllm6, secinllm7, secinllm8, secinllm9, secinllm10, secinllm11, secinllm12, secinllm13, secinllm14, secinllm15, secinllm16, secinllm17, secinllm18, secinllm19, secinllm20, secinllm21, secinllm22, secinllm23, secinllm24},
\emph{Vulnerability Detection}~\cite{vulndet1, vulndet2, vulndet3, vulndet4, vulndet5, vulndet6, vulndet7, vulndet8, vulndet9, vulndet10, vulndet11, vulndet12, vulndet13, vulndet14, vulndet15},
\emph{Generative AI Safety}~\cite{genai1, genai2, genai3, genai4, genai5, genai6, genai7, genai8, genai9, genai10, genai11, genai12, genai13, genai14}, 
\emph{Fuzzing}~\cite{fuzzing1, fuzzing2, fuzzing3, fuzzing4, fuzzing5, fuzzing6, fuzzing7, fuzzing8}, \emph{Secure Code Generation}~\cite{seccode1, seccode2, seccode3, seccode4}, 
and \emph{Vulnerability Repair}~\cite{vulnrep1, vulnrep2, vulnrep3, vulnrep4, vulnrep5, vulnrep6}. 
Figure~\ref{fig:paper-categories} provides an overview of the distribution of papers across these categories.

\subsection{Reviewing Methodology}\label{sec:review_methodology}

To guide the review process, we developed a detailed set of guidelines to ensure consistency across reviewers. 
These guidelines define how to identify each pitfall and provide instructions on interpreting relevant cues within the papers.
To refine the guidelines, we conducted a pilot review on a small subset of papers. This process helped clarify ambiguous cases and improve reviewer instructions. See Appendix~\ref{app:pitfall_guidelines} for the final version of our guidelines.

\label{sec:prevalence_assessment:labeling_scheme}
\boldpar{Labeling Scheme}
For each of the nine pitfalls, reviewers assigned one of several labels reflecting its applicability and presence.  
A pitfall is marked as \emph{Does not apply} (\colorcirclemaker{colorDoesNotApply}) if it is outside the scope for the particular paper, or as \emph{Unclear from text} (\colorcirclemaker{colorUnclearFromText}) when the paper lacks sufficient detail to support a clear judgment.
If the pitfall is applicable but not present, it is labeled as \emph{Not present}~(\colorcirclemaker{colorNotPresent}). We use \emph{Partly present}~(\colorcirclemaker{colorPartlyPresent}) to indicate that the pitfall affects only a part of the paper's methodology. A pitfall was marked as \emph{Present}~(\colorcirclemaker{colorPresent}) when it appears throughout the paper.
Finally, in case there is a high likelihood of a pitfall's presence but no explicit evidence, we use \emph{Likely present}~(\colorcirclemaker{colorLikelyPresent}). For example, take the case of potential data leakage where the evaluation dataset was publicly available online before the training-data cutoff of a model that has been trained on web-scale data. In such cases, we cannot prove that the evaluation data were included during training, but their presence is highly plausible due to timing and accessibility. 
This process is summarized in Figure~\ref{fig:decision_tree} in the Appendix.

It is important to note that some pitfalls may be difficult or impossible to avoid. Therefore, for any pitfall labeled as \emph{Likely present}, \emph{Partly present}, or \emph{Present}, we also annotated whether the presence of the pitfall was discussed in the paper. 
This distinction helps recognize good practice in transparently acknowledging potential limitations or contextualizing the relevance of a pitfall's presence. 
For example, \emph{\Plabelinaccuracy[]} (P\PIDlabelinaccuracy) may be unavoidable in large-scale evaluations where labeling is expensive, or  \emph{\Pdataleakage[]} (P\PIDdataleakage) may be difficult to address due to limited access to the original training data of proprietary models.

\boldpar{Reviewing Phases} 
The review was conducted by a team of fifteen researchers with backgrounds in security, machine learning, and software engineering. Every paper was assigned to exactly two reviewers, aiming to match papers with reviewers whose expertise aligns closely with the paper's topic. To ensure consistency and rigor throughout the review process, we structured the evaluation into three phases:
 
\italicpar{Phase 1: Individual Reviews} 
In the first phase, each reviewer independently assessed their assigned papers using the predefined guidelines. Reviewers were blinded to each other's assessments to ensure independent judgment.

\italicpar{Phase 2: Review Discussions}
In the second phase, reviewers discussed each paper to resolve differences from their initial assessments. Disagreements were clarified through 1-on-1 sessions or, when needed, group discussions (see Phase 3). For each pitfall, we recorded both the initial and final decisions.

\italicpar{Phase 3: Group Discussions} For cases where consensus could not be reached, we escalated the discussion to a larger group. These sessions included ten of the fifteen reviewers and served to resolve especially ambiguous or borderline cases.

\newcommand{\pitfalldescription}{\boldpar{Detailed Description}}
\renewcommand{\pitfalldescription}{\paragraph{Detailed Description}~}

\newcommand{\pitfallresults}{\boldpar{Results \& Implications}}
\renewcommand{\pitfallresults}{\paragraph{Results \& Implications}~}

\subsection{Pitfall Prevalence} \label{sec:prevalence_assessment:prevalence_results}

We now present the results of our prevalence study. As discussed above, we begin by examining each pitfall individually, summarizing how frequently it occurs across the reviewed papers and highlighting potential implications. In the following section (\S\ref{sec:prevalence_assessment:analysis}), we broaden the discussion to consider general patterns and shared challenges across pitfalls.

\begin{pitfall}[\PIDdatapoisoning]
A dataset used to train a model is collected from the internet without strategies to verify the integrity and safety of the data~\cite{poison_frogs}.
\end{pitfall}

\pitfallresults Although this pitfall was \emph{Present} in 4.2\% (3) and \emph{Likely present} in 23.6\% (17) of the papers, none acknowledged or discussed the risk of training on potentially poisoned internet data. This is especially concerning given recent work demonstrating that even minimal amounts of poison can compromise LLM behavior~\cite{mitre_poison, bowen_data_2024, wan_poisoning_2023}. In security-relevant settings, this can result in unsafe code suggestions or behavior misalignment triggered by crafted inputs.

\begin{pitfall}[\PIDlabelinaccuracy]
LLMs are used to annotate data with certain labels via classification or LLM-as-a-judge procedures without further validation of correctness.
\end{pitfall}

\pitfallresults This pitfall was \emph{Present} in 15.3\% (8) of the papers (five of which discussed the pitfall), \emph{Partly present} in 8.3\% (6) of the papers (four of which discussed it), and \emph{Likely present} in 1.4\% (1) of the papers. While not among the most common pitfalls overall, 60\% of relevant cases showed a relatively high awareness by discussing the issue. Nonetheless, unvalidated LLM-generated labels can distort results and lead to faulty conclusions. This is especially relevant in areas like jailbreak detection, where LLMs are often used as judges and external validation is critical to ensure~correctness.\smallskip

\begin{pitfall}[\PIDdataleakage]
An LLM is trained or fine-tuned with data that is normally not available in practice, or the training data is contaminated with possible test data~\cite{learning_from_data}.
\end{pitfall}

\pitfallresults \Pdataleakage[] is highly prevalent: 65.2\% (47) of the papers either contain it or are likely affected by it. Specifically, \Pdataleakage[] is \emph{Present} in 20.8\% (15) of the papers (of which seven explicitly discuss the pitfall), \emph{Partly present} in 6.9\% (5) of the papers (two of which discussed the pitfall), and \emph{Likely present} in another 37.5\% (27) of the papers (one of which discussed the pitfall). This raises concerns about inflated performance metrics, as models may memorize rather than generalize. Even state-of-the-art LLMs like \GPTmodel have suffered from such issues in tasks like code completion and vulnerability detection~\cite{dong-etal-2024-generalization}.

\begin{pitfall}[\PIDmodelcollapse]
An LLM is trained on data that is generated by other language models, risking an amplification of bias and degradation of data quality~\cite{shumailov_ai_2024, seddik_how_2024, guo_curious_2024}.
\end{pitfall}

\pitfallresults Although this pitfall was \emph{Present} in 13.9\% (10) of the papers, none of the papers in our study acknowledged or discussed the risk of using synthetic LLM-generated training data. Shumailov~\etal have demonstrated a decrease in model performance when trained on synthetic data, making the performance tied to the LLMs training data~\cite{shumailov_ai_2024}. Thus, training on LLM-generated data risks inheriting specific behaviors, biases, and errors~\cite{cloud2025subliminallearninglanguagemodels}. This can undermine generalization, particularly in tasks such as code generation, potentially distorting results and degrading performance.  Notably, for 37.5\% (27) of papers, this pitfall \emph{Does not apply}, as they do not involve model training.

\begin{pitfall}[\PIDspuriouscorrelations]
The LLM adapts to unrelated artifacts from the problem space instead of generalizing onto the actual task~\cite{risse_uncovering_2024}.
\end{pitfall}

\pitfallresults This pitfall appears in 31\% of papers: \emph{Present} in 5.6\% (4) of the papers (one of which discussed the pitfall), \emph{Likely present} in 23.6\% (17) of the papers, and \emph{Partly present} in 1.4\% (1) of the papers (which discussed it). Notably, it is discussed explicitly in only two papers. Due to the complexity of tasks and potentially black-box settings in which LLMs are deployed, Spurious Correlations often remain unidentified. This not only distorts experimental results but also leads to drawing false conclusions and using LLMs in scenarios for which they are not suited. In the security domain, for example, LLMs for vulnerability detection were unable to distinguish between vulnerable functions and the same functions that had been patched. The dependence on unrelated features led to top scores on benchmarks, even after removing the code structure itself~\cite{risse_top_2025}.

\begin{pitfall}[\PIDsmallcontext]
The LLM's context size is not large enough for its intended task, and the input needs to be truncated.
\end{pitfall}

\pitfallresults While not as prevalent as \Pdataleakage[], context limitations remain a notable pitfall: 33.3\% papers (24) were affected, either marked as \emph{Present} in 19.4\% of the papers (14, 6 of which discussed the pitfall), \emph{Partly Present} in 5.5\% of the papers (4, all of which discussed the pitfall), or \emph{Likely present} in 8.3\% of the papers (6, none of which discussed the pitfall). This is especially relevant in tasks like vulnerability detection~\cite{risse_top_2025} or prompt-based attacks~\cite{yu_jailbreaks}, where long inputs are common. A limited context size can cause models to overlook key information or distort evaluation results.

\begin{pitfall}[\PIDpromptsensitivity]
The prompt used to instruct the language models is fixed for all models and experiments, or is not expressive enough for the given task. This allows for prompt-based fluctuations in evaluations.
\end{pitfall}

\pitfallresults \Ppromptsensitivity[]\ affected 23 papers~(31.9\%) in our analysis: \emph{Present} in 18.1\% (13) of the papers  (3 of which discussed the pitfall), \emph{Likely present} in 6.9\% (5) of the papers, and \emph{Partly present} in 6.9\% (5) of the papers (2 of which discussed it). If prompt formatting is not adapted to the model, e.g., by using incorrect delimiters or generic instructions, this can reduce model performance or lead to misleading evaluations. In tasks such as jailbreak detection or alignment testing, vague prompts (e.g., ``You are a helpful AI assistant'') may underrepresent a model's true capabilities or vulnerabilities. As a result, evaluations might miss important failure modes or overstate alignment, leading to flawed comparisons across models~\cite{sclar2024quantifying}.

\begin{pitfall}[\PIDproxyfallacy]
Findings from specific LLMs are often inappropriately generalized to other, often larger and more capable models or even to entire classes of language models, without sufficient empirical validation.
\end{pitfall}

\pitfallresults \Pproxyfallacy[]\ is highly prevalent, appearing in 47.2\% (34) of the papers: \emph{Present} in 19.4\% (14) papers (3 of which discussed the pitfall), \emph{Partly present} in 26.4\% (19) of the papers (2 of which discussed it), and 1 paper is marked as \emph{Likely present}. Due to the distinct differences in behavior, claims about specific LLMs do not generalize to other models. Especially in security, where attacks and defenses hinge on specific model behavior, extrapolating results from smaller open-source models to larger proprietary ones is methodologically flawed and risks overstating generalizability.

\begin{pitfall}[\PIDmodelinfo]
The model details are insufficient for precise identification, preventing reproducibility (\eg, missing model ID, snapshot, commit ID, quantization level).
\end{pitfall}

\pitfallresults This is the most prevalent pitfall in our study: It was \emph{Present} in 73.6\% (53) of all papers, and \emph{Partly present} in 12.5\% (9). Notably, none of them acknowledged or discussed this issue, indicating a low level of community awareness. The lack of precise versioning details makes it nearly impossible to reproduce results reliably. Vendors frequently update LLMs to improve reliability or harden them against attacks, changes that can alter behavior without changing the model name. Similarly, quantization variants in open-source models can produce different outputs due to reduced precision. Microsoft highlighted this in the case of \modelID{Phi3mini4k}, where a major update significantly improved benchmarks; they published results for both versions and documented the relevant \mbox{commit~\cite{abdin_phi-3_2024, microsoftphi-3-mini-4k-instruct_2025}}.

\subsection{Main Findings} \label{sec:prevalence_assessment:analysis}

Every pitfall occurs in multiple papers, and every paper contains at least one pitfall that is either fully or partly present (\cf Figure~\ref{fig:pitfall-overview} in the Appendix for a side-by-side comparison of all pitfalls). The most common issues, present in over 20\% of papers, are \textit{\Pdataleakage[]\ (P\PIDdataleakage)}, \textit{\Psmallcontext[]\ (P\PIDsmallcontext)}, \textit{\Pmodelinfo[]\ (P\PIDmodelinfo)}, \textit{\Ppromptsensitivity[]\ (P\PIDpromptsensitivity)}, and \textit{\Pproxyfallacy[]\ (P\PIDproxyfallacy)}. We analyze the impact of four of these in \S\ref{sec:impact_analysis}.

Furthermore, for three pitfalls, namely \textit{\Pdataleakage[]\ (P\PIDdataleakage)}, \textit{\Pdatapoisoning[]\ (P\PIDdatapoisoning)}, and \textit{\Pspuriouscorrelations[]\ (P\PIDspuriouscorrelations)}, the \emph{Likely present} label applies to more than 20\% of papers, which means that the pitfall was probably present despite the absence of explicit evidence.

\result{\textbf{Finding.} Every paper in our study contains at least one pitfall that is either fully or partly present.}

\boldpar{Discussed vs. Not Discussed}
Overall, only 15.71\% of pitfalls rated at least \emph{Likely present} (\emph{Likely present}, \emph{Partly present}, or \emph{Present}) were discussed by the authors of the papers. The three most frequently discussed pitfalls were \textit{\Plabelinaccuracy[]\ (P\PIDlabelinaccuracy)} with 60.0\%, \textit{\Psmallcontext[]\ (P\PIDsmallcontext)} with 41.67\%, and \textit{\Pdataleakage[]\ (P\PIDdataleakage)} with 21.28\% of cases being explicitly addressed. In contrast, some pitfalls received no discussion at all: \textit{\Pmodelcollapse[]\ (P\PIDmodelcollapse)}, \textit{\Pdatapoisoning[]\ (P\PIDdatapoisoning)}, and \textit{\Pmodelinfo[]\ (P\PIDmodelinfo)}, indicating a lack of awareness of these issues.
\result{\textbf{Finding.} Only 15.71\% of pitfalls are discussed, with three pitfalls (\textit{\Pmodelcollapse\ (P\PIDmodelcollapse)}, \textit{\Pdatapoisoning\ (P\PIDdatapoisoning)}, and \textit{\Pmodelinfo\ (P\PIDmodelinfo)}) remaining entirely undiscussed across all papers in our study.}

\begin{figure}[!t]
    \centering
    \begin{tikzpicture}
  \begin{axis}[
    ybar,
    bar width=18pt,
    bar shift=0pt,
    axis x line=box,
    xtick pos=bottom,
    ymin=0, ymax=30,
    ylabel={Average Pitfall Presence (\%)},
    symbolic x coords={
      Vulnerability Repair,
      Vulnerability Detection,
      Security in LLM,
      GenAI Safety,
      Secure Code Generation,
      Fuzzing,
    },
    xtick=data,
    xticklabel style={rotate=40, anchor=east, align=right},
    enlarge x limits=0.1,
    ymajorgrids,
  ]

  \addplot[draw=none, forget plot] coordinates {
    (Vulnerability Repair,0)
    (Vulnerability Detection,0)
    (Security in LLM,0)
    (GenAI Safety,0)
    (Secure Code Generation,0)
    (Fuzzing,0)
  };

  % Each bar with its own color
    \addplot+[draw=black, color=colorLLMForSecurity!80] coordinates {(Vulnerability Repair,27.78)};
    \addplot+[draw=black, color=colorLLMForSecurity!80] coordinates {(Vulnerability Detection,23.70)};
    \addplot+[draw=black, color=colorSecurityInLLM] coordinates {(Security in LLM,19.91)};
    \addplot+[draw=black, color=colorSecurityInLLM] coordinates {(GenAI Safety,19.84)};
    \addplot+[draw=black, color=colorLLMForSecurity!80] coordinates {(Secure Code Generation,17.78)};
    \addplot+[draw=black, color=colorLLMForSecurity!80] coordinates {(Fuzzing,15.28)};

  \end{axis}
\end{tikzpicture}
    \caption{Percentage of pitfalls labeled as \emph{Present} (including cases where the pitfall was discussed), averaged over all pitfalls and papers within each topic.
    Light blue bars (\colorcirclemaker{colorLLMForSecurity!80}) represent research that uses LLMs to address security problems. Dark blue (\colorcirclemaker{colorSecurityInLLM}) bars correspond to research focused on the security and safety of LLMs~themselves.}
    \label{fig:percentages_per_topic}
\end{figure}
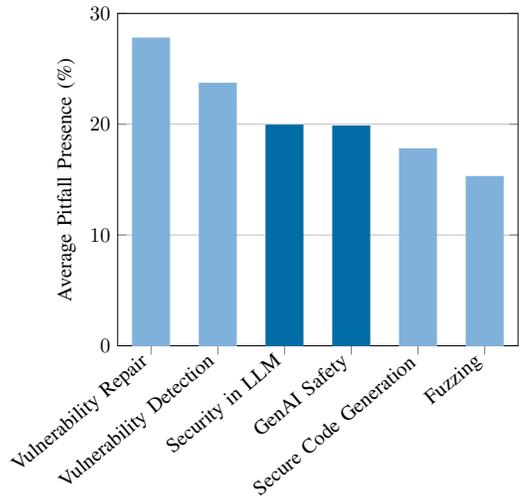

\boldpar{Topics} We further analyze the prevalence of pitfalls across the six research topics defined in our paper selection process. Figure~\ref{fig:percentages_per_topic} shows the average percentage of pitfalls per paper within each topic. This reflects the proportion of pitfalls labeled as \emph{Present}, including those where the pitfall was discussed.

The results show notable differences across topics. \emph{Vulnerability Repair} papers stand out with the highest average pitfall presence (27.78\%), followed by \emph{Vulnerability Detection} (23.70\%), indicating a high density of pitfalls in work that uses LLMs to find or fix software vulnerabilities. At the other end, \emph{Fuzzing} exhibits the lowest pitfall presence (15.28\%). \emph{Secure Code Generation} averages 17.78\%, suggesting this subfield is comparatively more robust with respect to the pitfalls we assessed. Research under \emph{Security in LLMs} (19.91\%) and \emph{GenAI Safety} (19.84\%) are close to 20\%.

\result{\textbf{Finding.} LLM-based research on \emph{Vulnerability Detection} and \emph{Vulnerability Repair} has the highest average number of pitfalls, whereas \emph{Secure Code Generation} and \emph{Fuzzing} papers tend to avoid such issues more consistently.}
\section{Impact Analysis} \label{sec:impact_analysis}

Having assessed the prevalence of pitfalls, we now examine their potential impact in greater detail. To this end, we focus on four of the most prevalent pitfalls identified in \S\ref{sec:prevalence_assessment}, each of which appears in at least 20\% of the papers. Specifically, we examine the impact of \Pmodelinfo[] and the \Pproxyfallacy[] (\S\ref{sec:no_detailed_model_version}), the influence of \Pdataleakage[] on experimental results (\S\ref{sec:impact_analysis:data_leakage}), and the limitations that \Psmallcontext[] imposes on model capabilities (\S\ref{sec:impact_analysis:context_size}). 

Together, these pitfalls illustrate the broader risks they pose to the integrity and reliability of LLM research. Another issue that appears in over 20\% of the reviewed papers is \Ppromptsensitivity[]. While important, \Ppromptsensitivity[] has already been extensively studied in prior work (e.g., \cite{sclar2024quantifying,cao2024on}). For this reason, we do not provide a separate analysis here.
Instead, we consider \Pmodelcollapse[] (\S\ref{sec:impact_analysis:model_collapse}), which has received little attention in the security context so far but has recently been shown to degrade model reliability in the text domain~\cite{shumailov_ai_2024}. We argue that understanding the potential implications for security research is necessary, especially as fine-tuned or retrained models become increasingly common.

\subsection{Model Ambiguity \& Surrogate Fallacy} \label{sec:no_detailed_model_version}
The most widespread issue in our analysis is \Pmodelinfo[], which is present in 73.6\% of all papers.
To examine the potential impact of this pitfall on experimental results, we first conduct a targeted evaluation of hate speech detection using state-of-the-art proprietary LLMs across multiple snapshots of the same base model. 
We then evaluate the robustness of proprietary models against prompt-based attacks across different model snapshots, as well as that of open-source models with varying levels of quantization. In this context, we also examine the impact of another common pitfall, the \Pproxyfallacy[], which appears in 47\% of the papers. 

\subsubsection{Hate Detection}

In the first experiment, we analyze the impact of (missing) LLM version or snapshot information for the example of hate speech detection. 
To this end, we revisit a recently proposed method to prevent waves of hateful comments that often build up for a particular topic, as happened, for instance, during the COVID-19 pandemic on $\mathbb{X}$ (formerly Twitter)~\cite{vishwamitra_moderating_2024}.
The core idea is to use an LLM to benefit from its enhanced reasoning capabilities, as detecting hate with all its nuances and newly introduced derogatory terms is a challenging task. Given a text, the LLM is instructed to answer a series of questions designed to guide its reasoning process.

\boldpar{Experimental Setup}
We re-implement the experiments from Vishwamitra~\etal~\cite{vishwamitra_moderating_2024} with their pre-labeled dataset of hateful and normal tweets from $\mathbb{X}$.
This dataset contains tweets from three polarizing topics termed hate waves (COVID-19 pandemic, US Capitol insurrection, and Russia's invasion of Ukraine). The COVID-19 pandemic is further divided into subtopics, from which we focus on the ``vaccine'' subtopic. The calibration and testing for each topic are done separately, with each hate wave divided into four quarters. For simplicity, we calibrate the detection method using examples from the first quarter and test on the second quarter of each wave.
The original publication stated the use of \model{\GPTfour}.
Thus, we test three \model{\GPTfour} snapshots available at \entity{OpenAI} at the time of writing.

The detection is done by extracting keywords using \model{KeyBERT} and testing the novelty with \entity{NLTK's} \model{WordNet}. Detailed specifications for all models can be found in Table~\ref{tab:pitfall6_llm_info_hate} of Appendix~\ref{app:supplementary_model_information:pitfall6}.
We then provide the text and the extracted targets and terms to the LLM \modelID{gpt-4.1-mini-2025-04-14} and instruct it to check which keywords are a target or derogatory term. Equipped with these new targets and terms, the prompt template from the original work~\cite{vishwamitra_moderating_2024} is used, which instructs an LLM to answer a series of questions to determine if a text is hateful or not. The extracted targets and terms are appended at the end of the prompt. Finally, we evaluate the LLM on the test set to decide on each posting and record this as its final decision.

\boldpar{Results}
Table~\ref{tab:wave_hate_results} demonstrates that the detection performance varies considerably across the different \model{\GPTmodel} snapshots. 

\begin{table}[ht]
    \centering
    \caption{Model Ambiguity in Hate Detection}
    \label{tab:wave_hate_results}
    \begin{tabular}{
        l
        S[table-format=2.2, table-space-text-post={\,\%}]<{\,\%}
        S[table-format=2.2, table-space-text-post={\,\%}]<{\,\%}
        S[table-format=2.2, table-space-text-post={\,\%}]<{\,\%}
    }
    \toprule
        \bfseries Model              & \mcb{Accuracy} & \mcb{Precision} & \mcb{Recall} \\
    \midrule
        \modelID{gpt-4-0613}         &         88.70  &          87.07  &       82.05 \\
        \modelID{gpt-4-0125-preview} &         77.16  &          95.52  &       41.03 \\
        \modelID{gpt-4.1-2025-04-14} &         82.45  &          88.07  &       61.54 \\
    \bottomrule
    \end{tabular}
\end{table}

An analysis of the LLM outputs suggests a shift in how different AI models classify hate speech. 
We exemplify this using the tweet ``Say it to my face, you tide pod eating, unvaccinated fuck'', which was labeled as \emph{hate} in the dataset. The \modelID{gpt-4-0613} model identifies derogatory and insulting words directed at anti-vaxxers and therefore categorizes the text as identity hate. In contrast, \modelID{gpt-4.1-2025-04-14} also recognizes the text's insulting content but determines that the language does not meet the threshold for identity hate, as it does not incite violence or hatred. The model considers the text as toxic only.
In general, this observation is also reflected in the precision and recall rates in Table~\ref{tab:wave_hate_results}. The models \modelID{gpt-4-0125-preview} and \modelID{gpt-4.1-2025-04-14} flag fewer tweets as hateful due to the higher threshold for hate, resulting in a significant drop in the recall rate. At the same time, the precision increases due to the stricter detection. 
While it is likely that such behavior can be adjusted through more tailored prompting, our aim with this experiment is to highlight the considerable, underestimated impact of the LLM~version. 

\newcommand{\mymidrule}{\cmidrule(lr){1-1}\cmidrule(lr){2-8}}
\renewcommand{\mymidrule}{\cmidrule(lr){1-8}}

\begin{table*}[ht]
  \centering
  \caption{Attack success rates (ASR) across different quantization levels, inference engines (if applicable), and model architectures. Each configuration is evaluated using \num{16}~prompt-based attacks, evenly distributed across a total of \num{1000}~trials.%
  % Experiment for the \emph{\Pmodelinfo} pitfall. Various Llama models---present in the analyzed literature---are evaluated on 16 different prompt-based attacks. Each model is evaluated in different quantization variants and, if available, from two different vendors (namely \entity{Ollama}~\cite{ollama_website} and \entity{TheBloke}~\cite{thebloke_2024}). Each variant is evaluated over \num{1000} attack trials in total.
  \label{tab:no_model_info_llama}
  }
  % resizebox is the root of all evil!
  %\resizebox{\linewidth}{!}{
  \begin{tabular}{
    l
    S[table-format=2.2, table-space-text-post={\,\%}]<{\,\%}
    S[table-format=2.2, table-space-text-post={\,\%}]<{\,\%}
    S[table-format=2.2, table-space-text-post={\,\%}]<{\,\%}
    S[table-format=2.2, table-space-text-post={\,\%}]<{\,\%}
    S[table-format=2.2, table-space-text-post={\,\%}]<{\,\%}
    S[table-format=2.2, table-space-text-post={\,\%}]<{\,\%}
    S[table-format=2.2, table-space-text-post={\,\%}]<{\,\%}
  }
  \toprule
    \bfseries Model         & \multicolumn{7}{c}{\bfseries Quantization} \\
                            \cmidrule(lr){2-8}
                            & \mc{2-bit} & \mc{3-bit} & \mc{4-bit} & \mc{5-bit} & \mc{6-bit} & \mc{8-bit} & \mc{No Quant.} \\
  \mymidrule
    CodeLlama 7b (Ollama)   &      69.52 &      40.95 &      15.49 &      18.61 &      15.69 &      14.29 &      18.21 \\
    CodeLlama 7b (TheBloke) &      67.35 &      70.41 &      58.16 &      62.24 &      61.22 &      63.27 &      19.39 \\
  \mymidrule
    Llama~2 7b (Ollama)     &       1.61 &       8.05 &       4.02 &       6.04 &       6.64 &       5.73 &       6.14 \\
    Llama~2 7b (TheBloke)   &      13.88 &      17.35 &      29.59 &      25.51 &       7.24 &       6.64 &      11.22 \\
  \mymidrule
    Llama~3.1 8b  (Ollama)  &      32.33 &      22.59 &      16.87 &      18.37 &      11.35 &      10.04 &      21.18 \\
    %Llama~3.1 8b (TheBloke) & x & x & x & x & x & x & x \\
  \bottomrule
  \end{tabular}
  %}
\end{table*}

\subsubsection{LLM Robustness}

We continue by evaluating the robustness of several LLMs against a suite of prompt-based attacks, including prompt injections, adversarial suffixes, and complex jailbreaks~\cite{evertz2024whispers}.
To do this, we adopt the \emph{secret-key game} introduced by Evertz et al.~\cite{evertz2024whispers}. In this setup, the model is given a \emph{secret key} in its system prompt, along with explicit instructions to keep the key confidential. The attack then attempts to elicit the secret from the model. If the model outputs the key as part of its response, the attack is considered successful. Each model is evaluated over 1,000 attack attempts, distributed evenly across 16 attack types. To minimize bias, system prompts are sampled from a set of 1,000 randomly selected variations.

We first examine proprietary models across multiple snapshots. As in the previous case, it is common to refer to these models by general handles (\eg, \model{\GPTfouro}), but these labels may internally refer to different model snapshots. As of the time of writing, we evaluate all available snapshots of OpenAI's \model{\GPTthreefiveturbo}, \model{\GPTfour}, \model{\GPTfourturbo}, and \model{\GPTfouro}. % July'25

In contrast to proprietary models, open-source models are typically not accessed through APIs and therefore do not expose versioned snapshots under the same identifier. Instead, on platforms such as \entity{Huggingface}\cite{huggingface} and \entity{Ollama}\cite{ollama_website}, they are provided in different variations of quantization, and older snapshots can be accessed via their commit history.
To complement the snapshot-based evaluation of proprietary models, we analyze how quantization affects robustness in open-source models. We test various quantization levels (2-bit through 8-bit) as well as the original unquantized snapshots. We evaluate three models (\model{CodeLlama-7b}\cite{roziere_codellama_2024}, \model{Llama-2-7b}\cite{touvron_llama_2023}, and \model{Llama3.1 8b}\cite{llama3_website}) and two different inference engines (\entity{Ollama}~\cite{ollama_website} and \entity{TheBloke}~\cite{thebloke_2024}). Detailed specifications for all models are listed in Table~\ref{tab:pitfall6_llm_info_robustness} in Appendix~\ref{app:supplementary_model_information:pitfall6}.

\boldpar{Results}
Table~\ref{tab:no_model_info_openai} presents the results for different snapshots of OpenAI's \model{\GPTmodel}. We observe that different snapshots lead to meaningful differences in robustness. In particular, \model{\GPTthreefiveturbo} exhibits significant variability across snapshots, though smaller but still notable differences are visible for the other models as well.
The effect of different quantization levels is summarized in Table~\ref{tab:no_model_info_llama}. Interestingly, models with stronger quantization, such as 2-bit and 3-bit, appear more vulnerable to prompt-based attacks. Additionally, the results indicate a difference between inference engines.
Note that these results not only demonstrate the potentially severe impact of the \Pmodelinfo[] pitfall, but also serve as an example for the \Pproxyfallacy[], since they show that findings do not necessarily transfer to other and particular larger models.

\begin{table}[ht]
  \centering
  \caption{%
  Attack success rate (ASR) across snapshots of current OpenAI models evaluated on \num{16}~different prompt-based attacks over \num{1000}~iterations per snapshot in total. \label{tab:no_model_info_openai}
  % Attack success rate (ASR) for the \emph{\Pmodelinfo} pitfall experiment. Different snapshots of current OpenAI GPT models are evaluated on \num{16}~different prompt-based attacks over a total of \num{1000}~iterations per snapshot. \label{tab:no_model_info_openai}
  }

  \begin{minipage}[t]{0.5\columnwidth}
  \setlength{\tabcolsep}{1pt}
  \centering
  \begin{tabular}{
    l
    S[table-format=2.2, table-space-text-post={\,\%}]<{\,\%}
  }
  \toprule
    \bfseries Model & \mcb{ASR} \\
  \midrule
     \bfseries \GPTthreefiveturbo \\
     $\hookrightarrow$ \modelID{gpt-3.5-turbo-1106}     &  4.63\\
     $\hookrightarrow$ \modelID{gpt-3.5-turbo-0125}     & 16.11\\  
  \midrule                                    
    \bfseries \GPTfourturbo\\                              
     $\hookrightarrow$ \modelID{gpt-4-1106-preview}     &  1.81\\
     $\hookrightarrow$ \modelID{gpt-4-0125-preview}     &  2.13\\
     $\hookrightarrow$ \modelID{gpt-4-turbo-2024-04-09} &  2.31\\
  \bottomrule
  \end{tabular}
  \end{minipage}~
  \begin{minipage}[t]{0.45\columnwidth}
  \setlength{\tabcolsep}{3pt}
  \begin{tabular}{
    l
    S[table-format=1.2, table-space-text-post={\,\%}]<{\,\%}
  }
  \toprule
    \bfseries Model & \mcb{ASR} \\    
  \midrule
    \bfseries \GPTfour \\
     $\hookrightarrow$ \modelID{gpt-4-0314}             &  9.56\\
     $\hookrightarrow$ \modelID{gpt-4-0613}             &  8.05\\
  \midrule                                                 
    \bfseries \GPTfouro \\                                 
     $\hookrightarrow$ \modelID{gpt-4o-2024-05-13}      &  1.01\\
     $\hookrightarrow$ \modelID{gpt-4o-2024-08-06}      &  0.10\\
     $\hookrightarrow$ \modelID{gpt-4o-2024-11-20}      &  0.20\\
  \bottomrule
  \end{tabular}
  \end{minipage}
\end{table}

Overall, these findings highlight the importance of reporting detailed model information, including snapshot identifiers and quantization parameters. Such details are essential for the reproducibility and interpretability of experimental results.
The observations are further supported by recent disclosures from Microsoft, showing that performance-relevant updates are being applied to their hosted models on \entity{Huggingface}. For example, the \modelID{Phi~3~mini~4k~instruct} model received an update together with a full disclosure about its new performance scores\footnote{\url{https://huggingface.co/microsoft/Phi-3-mini-4k-instruct}; see Release Notes and June 2024 Update}.

\subsection{Data Leakage} \label{sec:impact_analysis:data_leakage}
We continue with the analysis of another very prevalent pitfall, \Pdataleakage[], which occurs in 57\% of all papers we considered.
LLMs are typically trained on broad snapshots of public internet data collected up to a fixed cutoff date~\cite{brown_language_2020}. 
This practice introduces a critical concern: public benchmark datasets used to evaluate model performance may have been included in the training data, thereby undermining the validity of such evaluations.

To examine this issue, we focus on vulnerability detection as a representative task within LLM security research. It is widely studied, relies on well-known datasets, and is commonly used to benchmark the capabilities of code-oriented LLMs. We first conduct a controlled experiment designed to simulate leakage under lab conditions, followed by an empirical analysis of commercial models.

\subsubsection{Leakage in Lab Setting}
We consider three widely used vulnerability detection datasets: \dataset{Devign}~\cite{devign}, \dataset{DiverseVul}~\cite{diversevul}, and \dataset{PrimeVul}~\cite{primevul}. For each dataset, we construct splits for training (60\%), validation (20\%) and testing (20\%), and fine-tune a \model{CodeT5+}~\cite{wang_codet5plus_2023} model. Finally, we evaluate each model on the test split to obtain a baseline F1-score. 

To simulate leakage, we fine-tune five additional models per dataset for 10 epochs, each trained on an increasingly larger fraction of the test set added to the training data (20\%, 40\%, 60\%, 80\%, 100\%). We deliberately ignore the effects of limited context window size (512 tokens) and potential spurious correlations, as these factors apply equally across all data splits and do not affect the relative comparison. Full model configurations are provided in Appendix~\ref{app:supplementary_model_information:pitfall2}.

\begin{figure}[t]
    \centering
    \begin{tikzpicture}
\begin{axis}[
    width=12cm,
    height=8cm,
    xlabel={Leakage Percentage},
    ylabel={F1 Score (Full Test)},
    title={},%F1 Score vs Leakage Percentage},
    legend style={at={(0.5,-0.2)}, anchor=north, legend columns=3},
    grid=both,
    xmin=0, xmax=100,
    ymin=0.6, ymax=1.0,
    xtick={0,20,40,60,80,100},
    xticklabel={\pgfmathprintnumber{\tick}\%},
    ytick={0.6, 0.7, 0.8, 0.9, 1.0},
    every axis plot/.append style={thick}
]

\addplot[color=primarycolor, mark=*, mark size=3, thick] coordinates {
    (0, 0.616)
    (20, 0.708)
    (40, 0.780)
    (60, 0.845)
    (80, 0.913)
    (100, 0.973)
};
\addlegendentry{Devign}

\addplot[color=secondarycolor, mark=square*, mark size=3, thick] coordinates {
    (0, 0.600)
    (20, 0.671)
    (40, 0.728)
    (60, 0.785)
    (80, 0.839)
    (100, 0.819)
};
\addlegendentry{DiverseVul}

\addplot[color=tertiarycolor, mark=triangle*, mark size=4, thick] coordinates {
    (0, 0.617)
    (20, 0.683)
    (40, 0.749)
    (60, 0.789)
    (80, 0.855)
    (100, 0.890)
};
\addlegendentry{PrimeVul}

\end{axis}
\end{tikzpicture}
    \caption{ F1-scores of \model{CodeT5+} models fine-tuned on \dataset{Devign}, \dataset{DiverseVul}, and \dataset{PrimeVul} with varying amounts of test data (0–100\%) leaked into the training data.}
    \label{fig:case_study_leakage}
\end{figure}
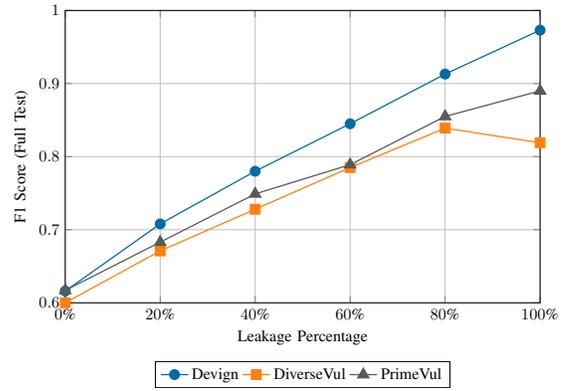

\boldpar{Results} The results are depicted in Figure~\ref{fig:case_study_leakage}. We observe a near-linear increase in F1-score with the degree of test set leakage. Even a modest leakage of 20\% leads to a gain of 0.08–0.11 over the baseline. 
This demonstrates that even partial leakage can significantly inflate evaluation metrics, leading to overly optimistic conclusions.

\subsubsection{Evaluation on Commercial LLMs}
Having observed the effects of leakage under controlled conditions, we now turn to the question of whether similar effects can be found in proprietary LLMs. The main challenge in this setting is the lack of transparency regarding the training data and procedures used for these models. 
To investigate this, we focus on the \dataset{PrimeVul}\cite{primevul} dataset. It consists of real-world vulnerabilities in public C/C++ projects, making it a plausible candidate for inclusion in the pre-training data of proprietary models. We evaluate four proprietary models from different vendors (\modelID{gpt-3.5-turbo-0125}, \modelID{gpt-4o-2024-08-06}, \modelID{DeepSeek-V3-0324}, \modelID{claude-3-5-haiku-20241022}) as well as three open-source models (\modelID{meta-llama-3-8b-instruct}, \modelID{qwen3-14b}, \modelID{qwen2.5-coder-14b}). For all evaluations, we set the temperature to 0. Detailed model information is available in Appendix~\ref{app:supplementary_model_information:pitfall2}. 

For each model, we conduct the following two experiments: (1) We sample 100 commits from the dataset, extract the original commit messages, and prompt the model with the project name, commit SHA, and the first half of the message, asking it to reconstruct the full commit message. (2) We repeat this process, using 100 sampled functions. We provide the same context, but instead of the commit message, we prompt the model with the first half of the function body and ask it to reconstruct the complete function.

\boldpar{Results} Surprisingly, none of the models successfully reconstructed a single commit message. Even allowing for fuzzy matching ($\geq 75\%$ word-level Jaccard similarity), all models yielded 0 out of 100 matches. The function completion task showed the same pattern: no exact or fuzzy matches across any of the models.
To rule out the possibility that the alignment procedures may suppress memorization, we fine-tune \modelID{gpt-3.5-turbo-0125} on 1,000 full commit messages (following the divergence attack from Carlini et al.~\cite{nasr_scalable_2023}). Despite this targeted fine-tuning, the model still failed to reproduce any of the original messages.

To investigate whether these commits may have been absent from the model's pre-training data, we searched all \dataset{Common Crawl} snapshots listed in the \model{\GPTmodel-3} paper~\cite{brown_language_2020} for any URLs related to the repositories used in \dataset{PrimeVul} (with wildcards). We found no matches. 
While we cannot rule out the use of private or proprietary data sources, the absence of \dataset{PrimeVul}-related content in \dataset{Common Crawl} suggests that \dataset{PrimeVul}-style data was likely not part of the pre-training mix.

These findings are somewhat unexpected, given that \dataset{PrimeVul} contains data from many of the most prominent and widely used open-source C/C++ projects on \entity{GitHub}. Nonetheless, we found no evidence that any of the tested models were trained on \dataset{PrimeVul} commits or functions. 

\boldpar{Discussion}
This finding is both surprising and encouraging. Despite the prominence of the underlying projects in \dataset{PrimeVul}, we find no evidence that the evaluated proprietary and open-source models were trained on these. 
While we cannot definitively rule out the possibility of data leakage, the results provide strong evidence that data leakage is unlikely in this case.
To support future evaluations and reduce the risk of leakage, we recommend the following: 
Ideally, test data should be drawn entirely from after the model's training cut-off date. In this case, it can be used with high confidence.
If the test data spans both before and after the cut-off, a pre/post probing analysis can help: a significant drop in performance on the post cut-off portion may indicate the absence of leakage. 
If the test data predates the cut-off or if timing is unclear, targeted memorization probes can be performed as done in our reconstruction experiments.
To aid this evaluation, Table~\ref{tab:train-align-data} in the Appendix summarizes publicly disclosed training data sources for several representative language models.

\subsection{Context Truncation} \label{sec:impact_analysis:context_size}
Next, we shift our focus to prompting, specifically the  issue of limited context windows. This pitfall appears in 28\% of the reviewed papers.
To this end, we examine how constrained context length can affect the reliability of performance measurements and potentially lead to misleading conclusions.
More specifically, our goal is to understand whether limited context during training or evaluation can prevent a model from detecting vulnerabilities, thereby producing misleading performance results.
For instance, \cveID{CVE-2014-2669}\footnote{\url{https://nvd.nist.gov/vuln/detail/CVE-2014-2669}; see Appendix~\ref{app:too_small_context_size}} illustrates a case where the line responsible for an integer overflow lies outside a 512-token context window; hence, a model would be unable to detect it since the full context is not available.

\boldpar{Experimental Setup}
Among the 15 papers on vulnerability detection included in our  study, we observe that eight papers use at least one model with a context size of 512 tokens or less, one paper uses a model with a 1024-token context size, and eight papers use at least one model with a context size larger than 2048 tokens.

To assess whether these context sizes are sufficient, we analyze function lengths in the datasets used in \S\ref{sec:impact_analysis:data_leakage}.
Using the \model{CodeT5}\cite{wang_codet5_2021} tokenizer, we compute the number of tokens for each function labeled as vulnerable. We then calculate the proportion of functions in each dataset that exceed context limits of 512, 1024, and 2048 tokens. This allows to quantify the proportion of functions that are partially truncated under typical context size assumptions. Further details on the tokenizer are provided in Appendix~\ref{app:supplementary_model_information:pitfall3}.

\boldpar{Results} We find that a substantial proportion of vulnerable functions exceed the context sizes used in papers on vulnerability detection in our study. As shown in Table~\ref{tab:token-cutoffs}, on average, 49.3\% of all vulnerable functions across \dataset{Devign}, \dataset{DiverseVul}, and \dataset{PrimeVul} contain more than 512 tokens when tokenized with the \model{CodeT5} tokenizer. At the 1024-token threshold, 29.1\% still exceed the limit, and 13.7\% surpass 2048 tokens.

These findings show that using limited context windows systematically truncates a large portion of vulnerable functions, potentially removing the very code that contains or explains the vulnerability. As such, evaluation under small context settings may fundamentally misrepresent a model's performance and lead to misleading conclusions about its ability to detect vulnerabilities. Notably, these percentages represent a lower bound on the issue; Risse et al.~\cite{risse_top_2025} show that vulnerability detection often depends on code outside the function itself, meaning even full-function context may be insufficient in many cases.

\begin{table}[ht]
  %\scriptsize
  \setlength{\tabcolsep}{2.5pt}
  \centering
  \caption{Proportion of functions labeled as \emph{vulnerable} whose \model{CodeT5}-tokenized length exceeds context-window sizes used in the papers on vulnerability detection identified by our pitfall study (\num{512}, \num{1024}, and \num{2048}~tokens).\label{tab:token-cutoffs}}
  \begin{tabular}{
    l@{\,}%
    S[table-format=6.0]
    S[table-format=5.0]@{~(}S[table-format=2.1, table-space-text-post={\%}]<{\%)}
    S[table-format=5.0]@{~(}S[table-format=2.1, table-space-text-post={\%}]<{\%)}
    S[table-format=5.0]@{~(}S[table-format=2.1, table-space-text-post={\%}]<{\%)}
  }
  \toprule
   \bfseries Dataset & \bfseries\#~Funcs & \multicolumn{6}{c}{\bfseries\# Tokens}  \\
                                           \cmidrule(lr){3-8}
                     &                   & \multicolumn{2}{c}{$>512$} & \multicolumn{2}{c}{$>1024$} & \multicolumn{2}{c}{$>2048$}  \\
  \midrule
    %BigVul           & 10895   & 4238 & 38.9   & 2299 & 21.1   & 1068 &  9.8 \\
    Devign           & 12460   & 5196 & 41.7   & 2642 & 21.2   &  984 &  7.9 \\
    DiverseVul       & 18945   & 9814 & 51.8   & 5835 & 30.8   & 2747 & 14.5 \\
    PrimeVul         &  6004   & 3897 & 64.9   & 2594 & 43.2   & 1357 & 22.6 \\
  \midrule
    \bfseries
    Average          & {--}    & \multicolumn{2}{c}{\bfseries\num{52.8}\%}
                               & \multicolumn{2}{c}{\bfseries\num{31.7}\%}
                               & \multicolumn{2}{c}{\bfseries\num{15.0}\%} \\
  \bottomrule
  \end{tabular}
\end{table}
% Lea stopped here

\subsection{Model Collapse}
\label{sec:impact_analysis:model_collapse}
\begin{figure}[t]
    \centering
    \includegraphics[width=\linewidth]{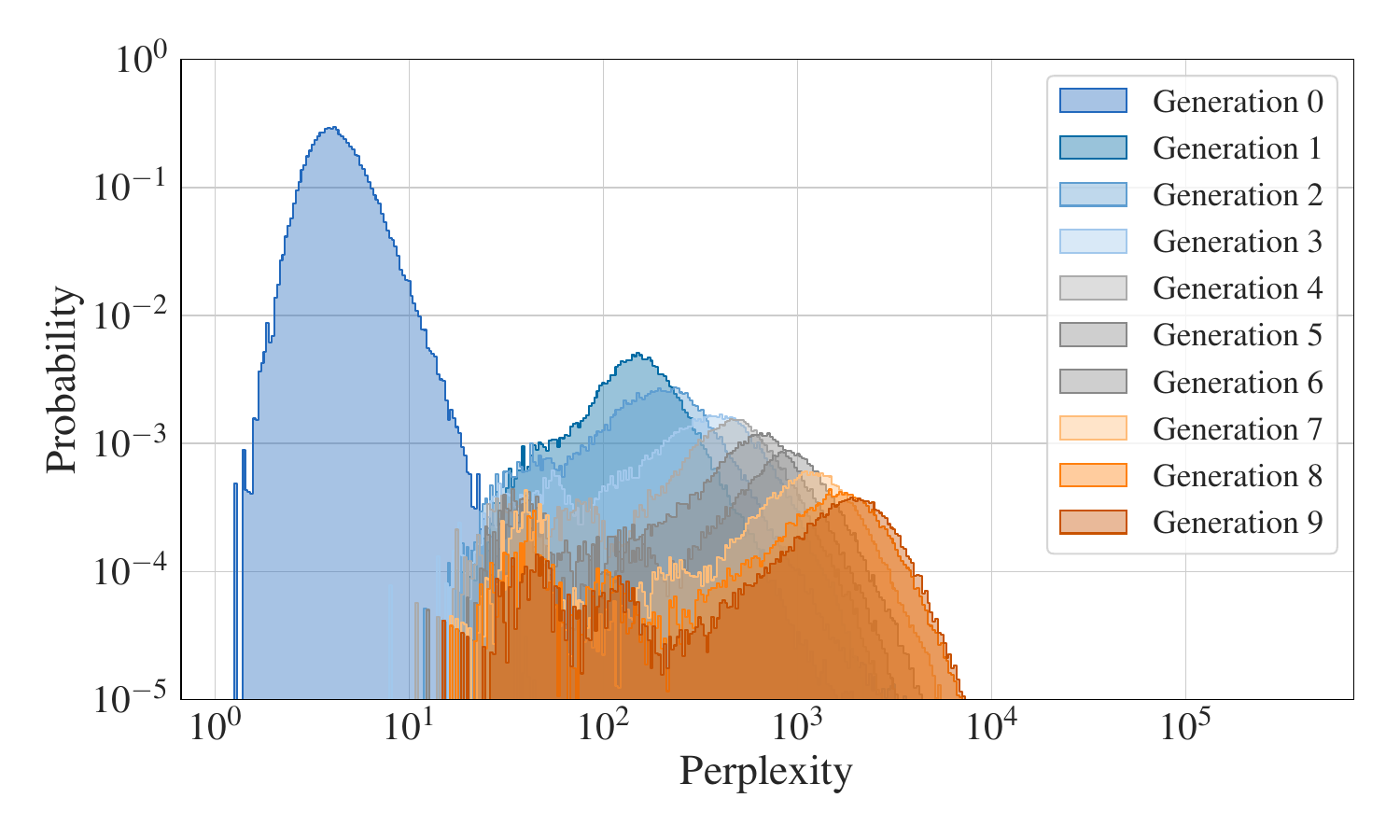}
    %\includestandalone[width=\linewidth]{figures/tex/GenerationPlot}
    %\includestandalone[width=\linewidth]{figures/tex/GenerationPlotFillCorner}
    \caption{%Histograms
    Distribution of perplexities for training samples generated across multiple model generations. Perplexity is measured using the original model fine-tuned on real data. Both the mean and variance of perplexity increase over generations, suggesting that models become less stable when repeatedly trained on synthetic data.
    % Distribution of perplexities of all individual training samples produced by models over multiple different generations and evaluated on the original model trained on the real dataset. Later generations are indicated by warmer colors.
    %The perplexities are right-shifting over the generations, indicating that the models become more unstable when trained on synthetic data several times.
    % Mean perplexity and variance increase with more generation, indicating that each model generation becomes more unstable when trained on synthetic data several times.
    % The higher perplexity is the result of samples that would never be produced by the original model hinting that the later generations accumulated errors from their ancestors.
    }
    \label{fig:perplexity_hist_bs128}
\end{figure}

Finally, we consider another important aspect of LLM training: the potential effects of model collapse, which we found to be present in 13.9\% of the papers.
Repeated training on synthetic data, particularly content generated by models, can gradually degrade model quality, amplify existing biases, and reduce output diversity~\cite{shumailov_ai_2024}.

\boldpar{Experimental Setup}
We build on the setup introduced by Shumailov~\etal~\cite{shumailov_ai_2024}, who studied model collapse in natural language generation.
Code generation is more challenging than generating natural language text, as source code is subject to stricter syntactic and semantic constraints. For this experiment, we use the \model{Qwen2.5-Coder-0.5B-Instruct}~\cite{qwen2, hui2024qwen2} model (full configuration details are provided in Table~\ref{tab:pitfall1_llm_info_collapse} in Appendix~\ref{app:supplementary_model_information:pitfall1}) and consider training sequences with a fixed length of 128 tokens in line with the setup from Shumailov~\etal~\cite{shumailov_ai_2024}.

We begin by fine-tuning the base model on the \dataset{self-oss-instruct-sc2-exec-filter-50k}~\cite{bigcodeself-oss-instruct-sc2-exec-filter-50k_2024, cassano_knowledge_2024} dataset, which contains 50,000 Python code samples. For each sequence, we predict up to 2,048 subsequent tokens to generate synthetic samples, ensuring that no code sample is truncated or incomplete. These synthetic outputs form a new dataset of equal sample size, which is then used to fine-tune the next model iteration. We repeat this process for ten generations, where each model is trained on data generated by its immediate predecessor. Although it cannot be ruled out that the base model is trained on (subsets) of this dataset, the experiment's primary focus is on the re-training on LLM-generated data. Hence, a possible data leakage would not affect the results.% Also, since this experiment serves as a proof of concept, we opt for a smaller model of the \model{Qwen2.5 Coder} family to reduce hardware and energy requirements and runtime.

\boldpar{Results}
Figure~\ref{fig:perplexity_hist_bs128} shows the distribution of perplexities across generations, measured by evaluating each generation's outputs using the fine-tuned base model.
\textit{Generation 0} refers to the fine-tuned base model evaluated on the unmodified training data. \textit{Generation 1} corresponds to the synthetic outputs from the first-generation model, and so on.

We observe a clear trend of degradation in model performance. With each generation, the distribution shifts to the right, indicating increased perplexity and suggesting that the base model becomes increasingly uncertain when predicting tokens in samples produced by later generations.
We also observe growing variance in perplexity across generations, reflecting greater instability in models trained on synthetic data, and indicating that errors accumulate over time. Furthermore, models trained primarily on synthetic outputs fail to generalize effectively.

These results extend prior concerns raised by Shumailov~\etal~\cite{shumailov_ai_2024} to the code domain. Our findings suggest that similar degradation can occur in code-related applications as well.
This trend is especially concerning as AI-powered coding assistants become more widespread and, thus, an increasing share of source code may be generated by models rather than written by humans.
From a security perspective, this trend also poses a serious risk. For example, when the underlying models are used for code generation, their reliability may degrade over time, potentially introducing subtle bugs or vulnerabilities. Similarly, when such models are used for, e.g., vulnerability detection, the degradation of the model may result in a lower detection rate.
\section{Recommendations} \label{sec:recommendations}
Our analysis reveals a concerning pattern: every reviewed paper suffers from at least one pitfall, and each pitfall appears in multiple papers. In the following, we propose recommendations that directly target the pitfalls while considering real-world constraints such as the use of proprietary systems, limited access to training data, and computational limitations.

We provide an extended discussion covering each pitfall individually in Appendix~\ref{app:recommendations_appendix}. In addition to this discussion, we offer a project website with guidelines to avoid pitfalls and a living appendix where all information is open for further contributions and kept up to date. The website is available at \url{https://llmpitfalls.org}.

\boldpar{R1: Transparency and Reproducibility}
Transparent reporting and reproducibility are crucial in research. 
Even if full reproducibility is not always feasible, especially with proprietary APIs, researchers should share enough information to enable others to interpret, replicate, or critically evaluate the findings.

\medskip
\noindent We recommend the following reporting practices:
\begin{itemize}
\item Specify the exact model used, including version identifier, access method (e.g., API or interface), and access date. For open-source models, include repository links, commit hashes, and any fine-tuning steps that were taken. Describe the evaluation pipeline in detail, including prompt format, quantization level, decoding settings, and any post-processing.
As shown in \S\ref{sec:no_detailed_model_version}, missing or vague model details can lead to substantial variation in outcomes and hinder reproducibility ($\longrightarrow$ \emph{\Pmodelinfo~(P\PIDmodelinfo)}).
\item If evaluation labels are generated by language models or if LLMs are used as judges, disclose this and assess label quality via manual review on a statistically meaningful subset. Report inter-annotator agreement where applicable ($\longrightarrow$ \emph{\Plabelinaccuracy\ (P\PIDlabelinaccuracy)}).
\item Avoid generalizing beyond what the evidence supports. Claims should be scoped to the specific models evaluated unless a diverse and representative set of models is tested and limitations are explicitly acknowledged ($\longrightarrow$~\emph{\Pproxyfallacy\ (P\PIDproxyfallacy)}).
\end{itemize}

\boldpar{R2: Understand and Communicate Data Quality}
Data plays a central role in both training and evaluation of language models. Whether data is synthetic, scraped, or manually labeled, its origin, quality, and security can significantly influence model behavior and reported results.

\medskip
\begin{minipage}{\linewidth}
\noindent To mitigate these risks, we recommend:
\begin{itemize}
\item When using synthetic or LLM-generated data for training, clearly report its proportion relative to real data and assess distributional differences. In iterative self-training setups, monitor for signs of model collapse or performance degradation. As demonstrated in \S\ref{sec:impact_analysis:model_collapse}, iterative training on synthetic data can lead to significant performance decline and instability over time ($\longrightarrow$ \emph{\Pmodelcollapse\ (P\PIDmodelcollapse)}).
\item For tasks with realistic poisoning threats, such as vulnerability detection or misinformation detection, explicitly assess whether poisoning is plausible in the training setup, particularly when relying on large-scale web data or proprietary models ($\longrightarrow$ \emph{\Pdatapoisoning\ (P\PIDdatapoisoning)}).

\item Acknowledge any risk of data leakage by checking whether evaluation data (especially answers or labels) was publicly available prior to the model's training cutoff date. Where cutoff dates are known, analyze performance on pre- vs. post-cutoff data and consider probing for memorization. We demonstrate the impacts of potential data leakage in \S\ref{sec:impact_analysis:data_leakage}. Whenever possible, prefer models for which information about training data composition is available (see Table~\ref{tab:train-align-data}). As a community, we should push for state-of-the-art models with transparent or public training data. While this is often not possible for proprietary models, ongoing regulatory efforts such as the EU AI Act’s transparency requirements for General-Purpose AI (GPAI) providers~\cite{eu_gpaimodel_template_2025} are expected to improve disclosure practices ($\longrightarrow$ \emph{\Pdataleakage\ (P\PIDdataleakage)}).
\end{itemize}
\end{minipage}

\boldpar{R3: Alignment between Models and Tasks}
Many evaluation failures stem not from the models themselves, but from a mismatch between the model constraints and the task requirements. Researchers should assess whether the models used are capable of handling the inputs, outputs, and task complexity involved and report limitations clearly.

\medskip
\noindent To that end:
\begin{itemize}
\item Report the model's maximum context window and analyze whether typical input sizes (including prompts) exceed this limit. If truncation occurs frequently, quantify how often and assess the potential impact on performance. Detailed evaluation of possible side effects of a too small context size is given in \S\ref{sec:impact_analysis:context_size} ($\longrightarrow$ \emph{\Psmallcontext\ (P\PIDsmallcontext)}).
\item Recognize that prompt design can meaningfully influence model behavior. When feasible, use established prompting techniques or conduct prompt variation experiments to gauge robustness. If prompt optimization is not applicable, explain why ($\longrightarrow$ \emph{\Ppromptsensitivity\ (P\PIDpromptsensitivity)}).
\item Evaluate whether the model may be exploiting spurious correlations. Use robustness checks (e.g., input perturbations, ablations) or feature attribution techniques to test whether the model relies on unintended features ($\longrightarrow$ \emph{\Pspuriouscorrelations\ (P\PIDspuriouscorrelations)}).
\end{itemize}
\section{Discussion} \label{sec:limitations}

Our study provides a systematic analysis of common pitfalls in LLM security research. However, like any such effort, it is not without limitations. In the following, we reflect on potential threats to validity and the generalizability of our findings, as well as the broader implications of the issues we uncovered. 

\boldpar{Addressing Inevitable Pitfalls} 
Some pitfalls in LLM research are difficult to avoid due to the inherent complexity and scale of current systems. The vast amount of data required to train LLMs introduces inherently challenging-to-manage risks. For example, \Pdatapoisoning[]\ attacks (P\PIDdatapoisoning), although realistic in principle, are challenging to detect and mitigate, especially when experiments rely on publicly available data. Similarly, the lack of transparency from LLM providers regarding training corpora increases the likelihood of \Pdataleakage[]\ (P\PIDdataleakage). In addition, the complexity and often opaque behavior of LLMs can lead to \Pspuriouscorrelations[]\ (P\PIDspuriouscorrelations) that are difficult to identify or explain. Although these issues cannot always be fully resolved (see \S\ref{sec:recommendations} for possible mitigation strategies), researchers must acknowledge, address, and ideally explicitly discuss such pitfalls when reporting their findings.

\boldpar{Coverage of Literature and Pitfalls} Our paper selection may not be exhaustive. Relevant work may have been missed due to limitations in our keyword filters or because it appeared outside the selected venues. To reduce this risk, we conducted an intentionally broad search across eight leading security and software engineering conferences, used an extensive keyword list, and applied a two-stage manual selection process. This resulted in 72 in-scope papers. While omissions remain possible, we believe they are unlikely to affect our qualitative conclusions.

Likewise, other pitfalls may exist beyond the nine we identified for our study. We iteratively refined our pitfall list and definitions during test reviews with a diverse reviewing team to limit this threat. We encourage future work to build on and extend this taxonomy.

\boldpar{Review Subjectivity} 
While some degree of subjectivity is unavoidable in qualitative reviews, we took several steps to improve consistency. Each paper was reviewed by two independent reviewers, who were blinded during the first phase of evaluation. We then conducted a structured discussion phase to reconcile disagreements.

To quantify inter-reviewer consistency, we computed Fleiss' $\kappa$ for each pitfall. Overall agreement was solid, with a mean $\kappa$ of $0.55$ across all pitfalls.
Out of 648 individual ratings (72 papers $\times$ 9 pitfalls), only 25 required resolution through discussion in the larger reviewer group. This corresponds to less than 5\% of all decisions. 
Most disagreements were minor and resolved through brief conversations between reviewers.
Full $\kappa$ values and qualitative interpretations are reported in Table~\ref{tab:kappa} in the Appendix.

\boldpar{Generalizability of the Impact Analysis}
Our experimental study focuses on four case studies and five selected pitfalls, covering key stages of the LLM pipeline. Thus, four pitfalls remain unexplored in the experimental analysis, and even for the five chosen ones, their impact may vary across tasks, datasets, and model families.
We emphasize that the impact analysis aims to demonstrate plausible mechanisms and concrete examples rather than provide generalizable effect sizes. As such, the results should be interpreted as illustrative lower bounds, not as comprehensive or definitive impact measures.

\section{Conclusion} \label{sec:conclusion}

In this work, we systematically reviewed 72 papers from leading security and software engineering conferences and identified nine recurring pitfalls in LLM security research. Every paper we considered contained at least one of these pitfalls, and each pitfall appeared across multiple studies. Our impact analysis shows how seemingly minor issues, such as \Pdataleakage[], \Psmallcontext[], or \Pmodelinfo[], can distort results, reduce reliability, and undermine reproducibility.
Together, these findings highlight the broader risks these pitfalls pose to the integrity of LLM security research and underscore the need for more rigorous, transparent, and reproducible practices.

Looking ahead, strengthening LLM security research involves applying the lessons we outline: reporting exact model identifiers, testing for leakage, using task-appropriate context sizes, verifying LLM-generated labels, and avoiding overgeneralization. Our guidelines are not meant to serve as a fixed checklist. However, by treating these pitfalls as design constraints rather than afterthoughts, the community can move from ``chasing shadows'' to conducting LLM security research that is reproducible, trustworthy, and grounded in evidence.
\section*{Ethical Considerations} \label{sec:ethics}
This work critically examines methodological patterns in LLM security research. To ensure transparency while minimizing the risk of reputational harm, we cite all sources transparently but report pitfalls in aggregate.
We deliberately refrain from assigning specific pitfalls to individual papers. Corresponding authors of the analyzed work are welcome to contact us for private discussion and clarification.

Our aim is constructive. Alongside our critique, we provide concrete, community-oriented guidelines to support more rigorous, transparent, and reproducible practices in LLM-focused security research.
\section*{Acknowledgments} \label{sec:acknowledgement}
This work was funded by the Deutsche Forschungsgemeinschaft (DFG, German Research Foundation) under Germany’s Excellence Strategy – EXC 2092 CASA – 390781972 and under the project ALISON (492020528), 
the Vienna Science and Technology Fund (WWTF) under the project BREADS (10.47379/VRG23011), 
the Helmholtz Association (HGF) within the topic "46.23 Engineering Secure Systems", 
the German Federal Ministry of Education and Research under the grant AIgenCY (16KIS2012) and SisWiss (16KIS2330), 
the European Research Council (ERC) under the consolidator grant MALFOY (101043410), and 
the LCIS center VW-Vorab-2025, ZN4704 11-76251-2055.
Additionally, Srishti Gupta was enrolled in the Italian National Doctorate on AI run by the Sapienza University of Rome in collaboration with the University of Cagliari during this project.
\bibliographystyle{IEEEtran}
\bibliography{strings, references, pitfalls_papers_references, misc}

\clearpage
\appendices

\begin{figure*}[ht]
    \centering
    \usetikzlibrary{shapes, arrows.meta, positioning}

\tikzstyle{decision} = [rectangle, draw, rounded corners, minimum width=3.4cm, minimum height=1cm, text centered]
\tikzstyle{leaf} = [decision, line width=2.5pt, text=white]
\tikzstyle{arrow} = [-{Latex}, thick, rounded corners]
\tikzstyle{textnode} = [text width=2.5cm, align=center]
\tikzset{drawf/.style={draw=#1,fill=#1}}

\newlength{\levelsep}
\setlength{\levelsep}{1.5cm}

\begin{tikzpicture}[node distance=\levelsep and 2.5cm]

% Top node
\node[decision] (applicable) {\textit{Applicable?}};

% Level 1
\node[decision, below=\levelsep of applicable, xshift=4.25cm] (present) {\textit{Present?}};
\node[leaf, below=\levelsep of applicable, drawf=colorDoesNotApply, xshift=-4.25cm, yshift=0.5mm] (notapply) {\textbf{Does not apply}};
\node[leaf, below=\levelsep of applicable, drawf=colorUnclearFromText, yshift=0.5mm] (unclear) {\textbf{Unclear from text}};

% Level 2
\node[leaf, below=\levelsep of present, xshift=-8.5cm, drawf=colorNotPresent] (notpresent) {\textbf{Not Present}};
\node[decision, below=\levelsep of present, xshift=-4.3cm, yshift=-26.0mm] (discussedLikely) {\textit{Discussed?}};
\node[decision, below=\levelsep of present, xshift=+7.7cm, yshift=-0.5mm] (partly) {\textit{Partly?}};

% Level 3
\node[decision, below=\levelsep of partly, xshift=-4cm] (discussedPartly) {\textit{Discussed?}};
\node[decision, below=\levelsep of partly, xshift=4cm] (discussedYes) {\textit{Discussed?}};

% Level 4
\node[leaf, below left=of discussedYes, xshift=4.0cm, drawf=colorPresent] (presentDiscussed) {\textbf{\shortstack{Present\\(but discussed)}}};
\node[leaf, below right=of discussedYes, xshift=-4.0cm, drawf=colorPresentBD] (presentLeaf) {\textbf{Present}};

\node[leaf, below left=of discussedPartly, xshift=4.0cm, drawf=colorPartlyPresent] (partlyDiscussed) {\shortstack[c]{\textbf{Partly present}\\\textbf{(but discussed)}}};
\node[leaf, below right=of discussedPartly, xshift=-4.0cm, drawf=colorPartlyPresentBD] (partlyLeaf) {\textbf{Partly present}};

\node[leaf, below left=of discussedLikely, xshift=4.0cm, drawf=colorLikelyPresent] (likelyDiscussed) {\shortstack[c]{\textbf{Likely present}\\\textbf{(but discussed)}}};
\node[leaf, below right=of discussedLikely, xshift=-4.0cm, drawf=colorLikelyPresentBD] (likelyLeaf) {\textbf{Likely present}};

% Arrows from top
\def\dy{0.5cm}

\draw[arrow] (applicable.south) -- ++(0,-\dy) -| node[above, near end, xshift=5mm, yshift=-2.4mm] {yes} (present.north);
\draw[arrow] (applicable.south) -- ++(0,-\dy) -| node[above, near end, xshift=8mm, yshift=-2mm] {unclear} (unclear.north);
\draw[arrow] (applicable.south) -- ++(0,-\dy) -| node[above, near end, xshift=-4.8mm, yshift=-2mm] {no} (notapply.north);

% Arrows from Present
\draw[arrow] (present.south) -- ++(0,-\dy) -| node[above, near end, xshift=-4mm, yshift=-2.0mm] {no} (notpresent.north);
\draw[arrow] (present.south) -- ++(0,-\dy) -| node[above, near end, xshift=6.5mm, yshift=-15.0mm] {likely} (discussedLikely.north);
\draw[arrow] (present.south) -- ++(0,-\dy) -| node[above, near end, xshift=6mm, yshift=-2.4mm] {yes} (partly.north);

% Arrows from DiscussedLikely
\draw[arrow] (discussedLikely.south) -- ++(0,-\dy) -| node[above, near end, xshift=4.3mm, yshift=-2mm] {no} (likelyLeaf.north);
\draw[arrow] (discussedLikely.south) -- ++(0,-\dy) -| node[above, near end, xshift=-5.2mm, yshift=-2.7mm] {yes} (likelyDiscussed.north);

% Arrows from Partly
\draw[arrow] (partly.south) -- ++(0,-\dy) -| node[above, near end, xshift=4.3mm, yshift=-2mm] {no} (discussedYes.north);
\draw[arrow] (partly.south) -- ++(0,-\dy) -| node[above, near end, xshift=-5.2mm, yshift=-2.7mm] {yes} (discussedPartly.north);

% Arrows from DiscussedYes
\draw[arrow] (discussedYes.south) -- ++(0,-\dy) -| node[above, near end, xshift=4.3mm, yshift=-2mm] {no} (presentLeaf.north);
\draw[arrow] (discussedYes.south) -- ++(0,-\dy) -| node[above, near end, xshift=-5.2mm, yshift=-2.7mm] {yes} (presentDiscussed.north);

% Arrows from DiscussedPartly
\draw[arrow] (discussedPartly.south) -- ++(0,-\dy) -| node[above, near end, xshift=4.3mm, yshift=-2mm] {no} (partlyLeaf.north);
\draw[arrow] (discussedPartly.south) -- ++(0,-\dy) -| node[above, near end, xshift=-5.2mm, yshift=-2.7mm] {yes} (partlyDiscussed.north);

\end{tikzpicture}
    \caption{Decision tree for deciding pitfall categories.}
    \label{fig:decision_tree}
\end{figure*}
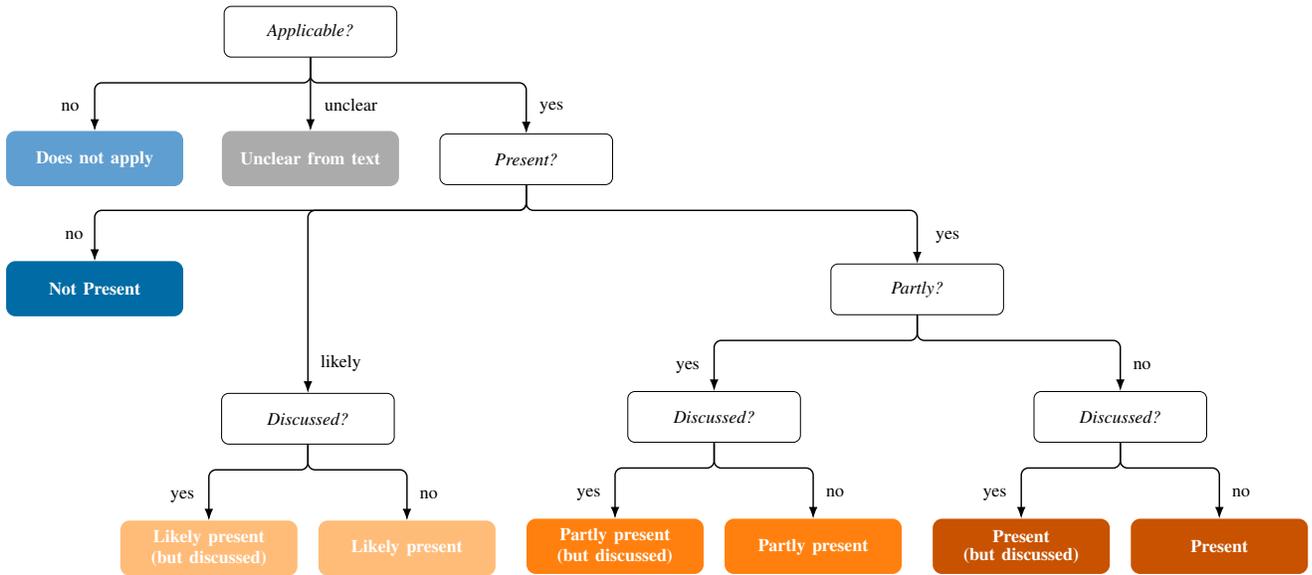

\section{Data Availability}\label{app:data_availability}
All code, datasets, and step-by-step reproduction instructions are available at: \url{https://github.com/dormant-neurons/llm-pitfalls}. A living appendix, with up-to-date information and guidelines to prevent pitfalls, is provided at \url{https://llmpitfalls.org}.

%%%%%%%%%%%%%%%%%%%%%%%%%%%%%%%%%%%%%%%%%%%%%%%%%%%%%%%%%%%%%%%%%%%%%%%
\section{Pitfall Guidelines}\label{app:pitfall_guidelines}

To ensure consistent evaluations across reviewers, we developed a set of clear guidelines for identifying potential pitfalls. These guidelines aim to promote clarity and reproducibility in the assessment of each case. For each pitfall, reviewers were instructed to consider the following questions:

\begin{enumerate}
    \item Is the pitfall \textbf{applicable} to the paper? Could it reasonably have influenced the results?
    \item Is the pitfall \textbf{present}? Is there clear evidence that it occurs in the paper, even if only to a limited extent?
    If there is strong indirect evidence or missing information that suggests the pitfall is probably present, then the pitfall is \textbf{likely present}.
    \item If present, is it \textbf{fully present} (affecting all results) or \textbf{partly present} (affecting only some results)?
    \item Is the pitfall \textbf{discussed} in the paper?
\end{enumerate}

A flowchart of the assessment process is depicted in Figure~\ref{fig:decision_tree}. In the following, we present the guidelines for each pitfall used in the prevalence study.

%%%%%%%%%%%%%%%%%%%%%%%%%%%%%%%%%%%%%%%%%%%%%%%%%%%%%%%%%%%%%%%%%%%%%%%
\begin{pitfall}[\PIDdatapoisoning][]\label{app:pitfall_guidelines:data_poisoning}
This pitfall applies if a dataset used to train a model is collected from the internet without strategies to verify the integrity and trustworthiness of the data (e.g., to check for poisoned examples).  
This applies even if the data is taken from a dataset published by a different paper, if no verification was performed there either.  
The pitfall applies only if there is training or fine-tuning in the paper; it does not apply otherwise.  
If data is collected by scraping third-party websites like GitHub or Stack Overflow without manual verification, the pitfall is \emph{likely present}.
\end{pitfall}

%%%%%%%%%%%%%%%%%%%%%%%%%%%%%%%%%%%%%%%%%%%%%%%%%%%%%%%%%%%%%%%%%%%%%%
\begin{pitfall}[\PIDlabelinaccuracy][]\label{app:pitfall_guidelines:label_inaccuracy}
This pitfall applies when LLMs are used to annotate data with labels via classification or "LLM-as-a-judge" procedures, without further validation of correctness.  
Check if the paper verifies the correctness of these labels or applies mitigation strategies.  
If LLM-as-a-judge is used to evaluate jailbreaks or attacks without further validation, this pitfall applies.
\end{pitfall}

%%%%%%%%%%%%%%%%%%%%%%%%%%%%%%%%%%%%%%%%%%%%%%%%%%%%%%%%%%%%%%%%%%%%%%%
\begin{pitfall}[\PIDdataleakage][]\label{app:pitfall_guidelines:data_leakage}
This pitfall refers to situations where information unavailable in real-world deployment is inadvertently included in the training data.  
Examples include:
\begin{itemize}
\item An LLM pre-trained or fine-tuned on data containing labels, metadata, or content from the test phase.
\item The model has access to future data during training that would not be available at inference time.
\end{itemize}
If only a part of the training datasets is affected, the pitfall is \emph{partly present}.  
The model must be directly affected by the data contamination. Otherwise, this pitfall is \emph{not present}.  
Since it is often hard to verify what data is present in the pretraining of an LLM, in most cases this pitfall will be \emph{likely present}.  
For example, \GPTmodel-2 was trained on Wikipedia, but we have no detailed sources for \model{\GPTmodel-4}, which makes it \emph{likely present}.  
If data is most likely present (such as Wikipedia), but we do not have clear proof, yet it is widely assumed in the community, then the pitfall is \emph{present}.
\end{pitfall}

%%%%%%%%%%%%%%%%%%%%%%%%%%%%%%%%%%%%%%%%%%%%%%%%%%%%%%%%%%%%%%%%%%%%%%%
\begin{pitfall}[\PIDmodelcollapse][]\label{app:pitfall_guidelines:model_collapse}
This pitfall applies if the model’s weights are influenced in any way (e.g., through finetuning) by synthetic data generated by an LLM.  
It also applies if external components, such as the tokenizer, are updated or trained with LLM-generated data.  
In the case of in-context learning, the pitfall does not apply, as there are no weight adjustments.  
Synthetic data refers to data produced as output by the same or a different LLM that is used for training or fine-tuning (see \S\ref{sec:impact_analysis:model_collapse} for rationale).
\end{pitfall}

%%%%%%%%%%%%%%%%%%%%%%%%%%%%%%%%%%%%%%%%%%%%%%%%%%%%%%%%%%%%%%%%%%%%%%%
\begin{pitfall}[\PIDspuriouscorrelations][]\label{app:pitfall_guidelines:spurious_correlations}
This pitfall applies when the LLM relies on spurious correlations or unrelated artifacts from the problem space, instead of learning to generalize to the underlying task.  
Check whether the model is capable enough for the chosen task.  
This pitfall also applies if reported performance varies considerably across models, suggesting the proposed approach is only effective for the specific model used.  
Look for evidence of explainability or interpretability analysis to determine what features the model relies on.  
Additionally, this pitfall applies if the same performance could be achieved with much simpler features (\eg, based on variable names or code formatting instead of semantics).
For example, a code vulnerability detector that performs well simply because vulnerable functions in the dataset tend to contain certain variable names, not because the model understands the logic.
\end{pitfall}

%%%%%%%%%%%%%%%%%%%%%%%%%%%%%%%%%%%%%%%%%%%%%%%%%%%%%%%%%%%%%%%%%%%%%%%
\begin{pitfall}[\PIDsmallcontext][]\label{app:pitfall_guidelines:context_size}
This pitfall applies if the LLM's context size is not large enough for the evaluation or its intended task, such that inputs need to be truncated.  
Check the model's maximum context length and the length of the used inputs. If not disclosed, estimate input size and convert to token counts using an online tool (\eg, LLM Token Counter).  
If the evaluation is affected by truncation, this pitfall applies.
\end{pitfall}

%%%%%%%%%%%%%%%%%%%%%%%%%%%%%%%%%%%%%%%%%%%%%%%%%%%%%%%%%%%%%%%%%%%%%%%
\begin{pitfall}[\PIDpromptsensitivity][]\label{app:pitfall_guidelines:prompt_sensitivity}
This pitfall applies if the prompt used to instruct the LLMs is either fixed across all models and experiments or lacks sufficient expressiveness for the specific task.  
The pitfall is considered \emph{present} if:
\begin{itemize}
    \item The study uses only a single prompt configuration (e.g., one prompt applied uniformly across all models) without justification or variation.
    \item Models are tested for robustness against adversarial inputs but are instructed using only generic prompts such as ``You are a helpful AI assistant.''
\end{itemize}
If the authors do not disclose how the prompting is designed and it appears generalized for all models, the pitfall is \emph{likely present}.  
If prompt variation is mentioned, but mitigation is insufficient or superficial, then the pitfall is \emph{present (but discussed)}.  
This pitfall does not apply to standard Machine Learning classification tasks (e.g., mapping code to labels using BERT) where prompts are not part of the input pipeline.
\end{pitfall}

%%%%%%%%%%%%%%%%%%%%%%%%%%%%%%%%%%%%%%%%%%%%%%%%%%%%%%%%%%%%%%%%%%%%%%%
\begin{pitfall}[\PIDproxyfallacy][]\label{app:pitfall_guidelines:proxy_surrogate_fallacy}
This pitfall applies when findings using specific LLMs are inappropriately generalized to other, sometimes larger, models or even to entire classes of language models, without sufficient empirical validation.  
This includes generalizing to different, untested quantization methods or different access methods (API vs. web), if evaluations were only done on one.  
The authors must make explicit claims for this pitfall to apply; do not mark as present for vague implications.  
For example, if an attack is tested on a small open-source \model{Llama-8b} model but the paper claims applicability to much larger or proprietary models like \model{\GPTmodel}, this pitfall applies.  
If claims are very vague (e.g., "LLMs cannot decipher obfuscated code from this framework"), mark as \emph{partly present}.
\end{pitfall}

%%%%%%%%%%%%%%%%%%%%%%%%%%%%%%%%%%%%%%%%%%%%%%%%%%%%%%%%%%%%%%%%%%%%%%%
\begin{pitfall}[\PIDmodelinfo][]\label{app:pitfall_guidelines:no_model_information}
This pitfall applies when model details are insufficient for precise identification, preventing reproducibility.  
Missing details can include:
\begin{itemize}
    \item Model IDs (e.g., mentioning only \model{\GPTfour} instead of a specific version like \modelID{gpt-4o-2024-11-20})
    \item Snapshots (for hosted models)
    \item Commit IDs (for local models, e.g., on \entity{Huggingface}~\cite{wolf_huggingfaces_2020} or \entity{Ollama}~\cite{ollama_website})
    \item Quantization level
\end{itemize}
If even one of these is missing, the pitfall is \emph{present} (not partly present).  
\emph{Partly present} is only used if the information is missing for only some of the models used (e.g., 1 out of 3).  
If using a hosted model instead of an open-source one, the paper must specify whether experiments are conducted via the API or the web interface, as these may differ in content moderation, system prompts, or other hidden context.  
If it is possible to reproduce the model version unambiguously (e.g., only one commit existed at the time of publication), then this pitfall does not apply.
\end{pitfall}

%%%%%%%%%%%%%%%%%%%%%%%%%%%%%%%%%%%%%%%%%%%%%%%%%%%%%%%%%%%%%%%%%%%%%%

\section{Recommendations} \label{app:recommendations_appendix}

Below is an extended version of the recommendation from \S\ref{sec:recommendations} aimed at helping researchers avoid the identified pitfalls in future work.

\begin{pitfall}[\PIDdatapoisoning][] To address the risk of data poisoning, researchers should first assess whether this threat is relevant to their specific task and data modality. For instance, in domains like vulnerability detection, data poisoning is particularly relevant, as adversaries may deliberately inject subtly hidden vulnerabilities into open-source repositories. If data poisoning is deemed relevant, the next step is to evaluate whether it is plausible in the given setup. For example, it could be plausible when using proprietary models or large-scale scraped datasets, where the training data is not fully transparent.

If data poisoning is both relevant and possible, researchers should explicitly acknowledge this risk in their paper. Following our prevalence assessment study, this corresponds to the \emph{Likely present (discussed)} label. While the ideal scenario would involve verifying the absence of poisoning in the training data, such guarantees are often unrealistic at scale.
\end{pitfall}

\begin{pitfall}[\PIDlabelinaccuracy][] When evaluation datasets include labels generated by LLMs, or when LLMs are used as judges for model outputs, researchers need to clearly disclose this and acknowledge the risk of inaccurate labels, which can lead to misleading performance evaluations. The optimal approach to mitigate this risk is to manually verify all such labels to ensure correctness. However, if a full audit is impractical due to scale, researchers should at least conduct a manual analysis of a statistically meaningful subset of the labels. This process should involve multiple annotators to minimize individual bias and should report inter-annotator agreement, along with confidence intervals, to assess reliability. 

While these precautions are essential when LLM-generated labels are used for evaluation, less care may be needed if high-quality, real labels are used for evaluation, and LLM-generated labels are only employed during fine-tuning or pre-training.
\end{pitfall}

\begin{pitfall}[\PIDdataleakage][]
Ideally, researchers should use open-source models with fully known training data and ensure there is no overlap with evaluation datasets. However, this is often not feasible, and there may be valid reasons to prefer proprietary models, such as access to stronger performance or specific capabilities.

To mitigate data leakage for such cases, the first step can be to identify the training data cutoff dates of the models under evaluation. For some proprietary models, this information is publicly available (e.g., for \entity{OpenAI} \model{o3}\footnote{\url{https://platform.openai.com/docs/models/o3}} or \entity{Anthropic} models\footnote{\url{https://docs.anthropic.com/en/docs/about-claude/models/overview}}). Researchers should then determine whether any portion of their evaluation dataset, especially labels or answers, was publicly accessible before this cutoff. 

However, as shown in our analysis (\S\ref{sec:impact_analysis:data_leakage}), relying solely on the cutoff date is insufficient. When feasible, researchers should probe for potential overlap, for example, by employing completion-style prompting or by comparing performance on data released before versus after the cutoff. Additionally, researchers can simulate leakage by intentionally inserting evaluation data into the training set (e.g., via fine-tuning) to estimate the resulting performance gains.
\end{pitfall}

\begin{pitfall}[\PIDmodelcollapse][] Using LLM-generated synthetic data can be a practical choice when real-world data is expensive or scarce. Rather than discouraging synthetic data altogether, researchers should critically assess its impact, particularly in terms of introducing biases and degrading model output quality---for instance, through iterative self-training, where an LLM is repeatedly trained on its own outputs, as demonstrated in Section~\S\ref{sec:impact_analysis:model_collapse}.

At a minimum, authors should clearly report the proportion of synthetic versus real data and analyze the differences between them. This is especially important in iterative training loops, where compounding effects can lead to model collapse by amplifying errors or reinforcing biases present in synthetic samples.
\end{pitfall}

\begin{pitfall}[\PIDspuriouscorrelations][] Exploiting spurious correlations is a fundamental issue in LLM behavior~\cite{2022_pavel_advances}. One approach to guard against this pitfall is robustness testing: by systematically perturbing input features, either those believed to be causally relevant or those suspected to be spurious, researchers can observe how model predictions change. If small changes to irrelevant input features (e.g., variable names, formatting, or unrelated context) significantly affect performance, this may indicate reliance on spurious features. Another approach is explainability techniques, such as feature attribution methods, which can help uncover which parts of the input the model focuses on during inference. 

Ultimately, researchers should attempt to falsify their own hypotheses and performance claims as much as possible. This includes conducting ablation studies and counterfactual evaluations aimed at verifying whether the model truly relies on features relevant to the problem it is supposed to solve.
\end{pitfall}

\begin{pitfall}[\PIDsmallcontext][] To mitigate issues caused by insufficient context size, researchers should begin by clearly stating the maximum context window of the model they are using. Next, they should assess whether this context window is adequate for the task at hand. One way to achieve this is tokenizing a representative sample of inputs, including the full prompt, and checking whether the total input size exceeds the model’s context limit, as demonstrated in our impact analysis in Section~\S\ref{sec:impact_analysis:context_size}. If a substantial portion of the input is truncated due to context size constraints, researchers should consider switching to a model with a larger context window. 

If this is not feasible, e.g., due to computational or financial limitations, they should, at a minimum, report what fraction of their inputs exceeds the context limit and discuss the implications. 

Additionally, researchers may analyze how model performance varies with input token length, which can help identify degradation patterns.
\end{pitfall}

\begin{pitfall}[\PIDpromptsensitivity][] To address the pitfall of prompt sensitivity, researchers should ideally optimize prompts for every specific task-model pair in their study. In practice, this means beginning with prompt design guidelines for the task at hand \cite{Chen2025Unleashing} and, when possible, leveraging prompt optimization techniques from prior literature \cite{pryzant-etal-2023-automatic}. 

\begin{figure*}[!t]
  \centering
  \setlength{\tabcolsep}{0pt}          
  \renewcommand{\arraystretch}{1.4}
  \newcommand{\presenceRaise}[1]{\raisebox{0\height}{#1}}

  % BEGIN: vertical lines for percentages
  \newcommand{\lineCount}{11}
  \newcommand{\leftMargin}{-4.125cm}
  \newcommand{\height}{11.7cm}
  \newcommand{\bottomMargin}{4.4cm}
  \newcommand{\lineSpacing}{1.328cm}
  \newcommand{\lineColor}{gray!20}
  \newcommand{\lineThickness}{0.3pt}

% das hier ist damit die Prozentzahlen rechts nicht über den Rand ragen
\resizebox{0.85\textwidth}{!}{%

  \begin{tikzpicture}[remember picture, overlay]
    \coordinate (tabAnchor) at ($(current page.center) + (0,0)$);
  \end{tikzpicture}
  
  \begin{tikzpicture}[remember picture, overlay]
    \foreach \i in {0,...,10} {
      \draw[\lineColor, line width=\lineThickness]
        ($ (tabAnchor) + (\leftMargin + \i*\lineSpacing + 0.525*\lineSpacing, \bottomMargin+0.95cm) $) --
        ($ (tabAnchor) + (\leftMargin + \i*\lineSpacing + 0.525*\lineSpacing, \height+0.9cm) $);
    }
  \end{tikzpicture}
  % END: vertical lines for percentages

  \begin{tabular}{r@{}l@{}}
    \Pdatapoisoning[] (P1)      & \presenceRaise{\presenceLine{3}{0}{1}{0}{17}{0}{30}{1}{20}} \\
    \Plabelinaccuracy[] (P2)    & \presenceRaise{\presenceLine{3}{5}{2}{4}{1}{0}{6}{0}{51}}   \\
    \Pdataleakage[] (P3)        & \presenceRaise{\presenceLine{8}{7}{3}{2}{26}{1}{5}{5}{15}}  \\
    \Pmodelcollapse[] (P4)      & \presenceRaise{\presenceLine{10}{0}{2}{0}{1}{0}{27}{1}{31}} \\
    \Pspuriouscorrelations[] (P5)
                             & \presenceRaise{\presenceLine{3}{1}{0}{1}{17}{0}{6}{8}{36}}  \\
    
    \Psmallcontext[] (P6)  & \presenceRaise{\presenceLine{8}{6}{0}{4}{6}{0}{3}{11}{34}}  \\
    \Ppromptsensitivity[] (P7)  & \presenceRaise{\presenceLine{10}{3}{3}{2}{5}{0}{13}{1}{35}} \\
    \Pproxyfallacy[] (P8)   & \presenceRaise{\presenceLine{11}{3}{17}{2}{1}{0}{0}{1}{37}} \\
    \Pmodelinfo[] (P9)   & \presenceRaise{\presenceLine{53}{0}{9}{0}{0}{0}{1}{0}{9}}   \\

    \noalign{\vskip 1em}
                             & \makebox[0pt][l]{\hspace*{-1.5mm}\presenceXAxis{72}} \\ \\
    \end{tabular}

% Ende von der Resizebox
}%
\vspace{-1.5em}
  \caption{Overview of the nine pitfalls across the 72 reviewed papers.
           Squares encode whether a pitfall in a paper is
            \emph{Present} (\colorcirclemaker{colorPresent}),
            \emph{Partly Present} (\colorcirclemaker{colorPartlyPresent}),
            \emph{Likely Present} (\colorcirclemaker{colorLikelyPresent}),
            \emph{Unclear from Text} (\colorcirclemaker{colorUnclearFromText}),
            \emph{Does not Apply} (\colorcirclemaker{colorDoesNotApply}), or
            \emph{Not Present} (\colorcirclemaker{colorNotPresent}).
           }
           
  \label{fig:pitfall-overview}
\end{figure*}

Even if full optimization is infeasible, researchers should conduct post-hoc prompt variation experiments to assess how changes in prompt phrasing influence model performance.

At a minimum, authors should clearly document how their prompts were constructed and explain the reasoning behind their design choices. If prompt optimization is not applicable, such as in cases where models are used in a fixed classification setup (e.g., mapping code to predicted labels), this should be explicitly stated to clarify the applicability of prompt sensitivity in the evaluation.
\end{pitfall}

\begin{pitfall}[\PIDproxyfallacy][] To avoid the proxy or surrogate fallacy, researchers should only make claims that are directly supported by the evidence presented in their paper. Specifically, they should refrain from drawing broad conclusions about entire model classes (e.g., \entity{LLMs}, \entity{Llama} models, or \model{\GPTmodel}) models based on results from a limited set of individual models. Instead, claims should be made at the level of the specific models evaluated, using precise identifiers as discussed in P\PIDproxyfallacy. 

If researchers wish to make broader claims about a model class, they must ensure that their evaluation includes a sufficiently diverse and representative sample of models within that class. Additionally, they should explicitly acknowledge that newer or differently configured models may behave differently and that their conclusions may not be generalizable.
\end{pitfall}

\begin{pitfall}[\PIDmodelinfo][] The ideal scenario is full reproducibility. Authors should release scripts or notebooks that enable end-to-end reproduction of their experiments, including data preprocessing, model interaction, and evaluation steps. However, full reproducibility may not always be possible due to factors such as proprietary models or dynamic APIs. In such cases, it is essential that researchers provide enough detail to ensure the exact models and configuration used can still be determined.

For proprietary models, researchers should report the exact model identifier (e.g., \modelID{o4-mini-2025-04-16}), the access method (e.g., web interface or API), and the date of access. This is particularly important as proprietary models may evolve over time or become unavailable. 

For open-source models, researchers should include precise information such as the model name, repository URL, commit hash, quantization level, and any fine-tuning steps applied. In both cases, researchers should be transparent about any parts of their pipeline that are not reproducible and explicitly acknowledge these limitations.
\end{pitfall}

\vspace{5em}
\section{Prevalence Assessment} \label{app:prevalence_assessment}
\begin{table}[ht]
\centering
\caption{Inter-reviewer agreement for each pitfall (Fleiss’ $\kappa$) and its Landis–Koch interpretation~\cite{landis_the_1977}.}
\label{tab:kappa}
\begin{tabular}{@{}lcc@{}}
\toprule
\textbf{Pitfall}                       & \textbf{$\kappa$} & \textbf{Interpretation}     \\
\midrule
\Pdatapoisoning[]\ (P1)                   & 0.52              & Moderate                    \\
\Plabelinaccuracy[]\ (P2)                 & 0.72              & Substantial                 \\
\Pdataleakage[]\ (P3)                     & 0.28              & Fair                        \\
\Pmodelcollapse[]\ (P4)                   & 0.43              & Moderate                    \\
\Pspuriouscorrelations[]\ (P5)            & 0.45              & Moderate                    \\
\Psmallcontext[]\ (P6)               & 0.39              & Fair                        \\
\Ppromptsensitivity[]\ (P7)               & 0.77              & Substantial                 \\
\Pproxyfallacy[]\ (P8)                & 0.27              & Fair                        \\
\Pmodelinfo[]\ (P9)                & 0.69              & Substantial                 \\
\bottomrule
\end{tabular}
\end{table}

%%%%%%%%%%%%%%%%%%%%%%%%%%%%%%%%%%%%%%%%%%%%%%%%%%%%%%%%%%%%%%%%%%%%%%%
\onecolumn

\section{Supplementary Model Information} \label{app:supplementary_model_information}
\colorlet{bgcolor}{gray!10!white}

\subsection{Model Ambiguity Pitfall} 

\boldpar{Hate speech experiment} The model specifics are listed in Table~\ref{tab:pitfall6_llm_info_hate}. The experiment was conducted using the default \model{\GPTmodel} parameters with $\texttt{temperature}=1.0$, its default unspecified system prompt, and a context size of $8,192$ tokens.

\begin{table*}[!h]
    \caption{Detailed information for the LLM used in the \emph{hate speech} experiment.}
    \label{tab:pitfall6_llm_info_hate}
    \centering
    \begin{tabular}{
        l
        cccc
    }
    \toprule
         \textbf{Model} & \textbf{Commit ID (w/ link) / Snapshot} & \textbf{Quantization} & \textbf{Access Type} & \textbf{Access Date} \\
    \midrule
         \model{\GPTfour} & \modelID{gpt-4-0613}, \modelID{gpt-4-0125-preview} and \modelID{gpt-4.1-2025-04-14} & not disclosed by provider & API & July 31st 2025\\
         \rowcolor{bgcolor}
         \model{KeyBert} & \modelID{all-MiniLM-L6-v2} & does not use quantization & locally hosted & July 31st 2025 \\
         \model{NLTK's WordNet} & \modelID{omw-en:1.4} & does not use quantization & locally hosted & July 31st 2025\\
    \bottomrule
    \end{tabular}
\end{table*}

\noindent The prompt format is shown in Figure~\ref{fig:hatespeech-prompt}.

\begin{figure}[!h]
\centering
\resizebox{0.55\linewidth}{!}{%
\begin{tcolorbox}[colback=gray!5!white, colframe=gray!75!black, title=Prompt Format for Hate Speech Experiment]
% \scriptsize
\ttfamily
        Consider the following comment, and carefully answer the questions in each step to conclude whether it is hate speech or not: 
        comment:"\{text\}"\\
\\
        Let's think of it in the following steps:\\
        Q1A: Which of the following identities are mentioned in the text?\\
        'race', 'nationality', 'age', 'political', 'religion', 'disability', '(anti-)masker', '(anti-)vaxxer'\\
        Q1B: Are there any individuals mentioned explicitly by their name?\\
        Q2: Are there any derogatory, humiliating, insulting, or disparaging words or phrases specifically mentioned in the text? (Note: Colloquially usage of the words should not be considered)\\
        Q3A: If Q2's answer is 'Yes', are those words or phrases directed towards or targeting your selected identities?\\
        Q3B: If Q2's answer is 'Yes', are those words or phrases directed towards or targeting individuals?\\
        Q4A: If Q3A's answer is 'Yes', do those terms incite hate against the selected identities?\\
        Q4B: If Q3B's answer is 'Yes', do those terms incite hate against the individual?\\
        Q5A: If Q4A's answer is 'Yes', the comment can be concluded as identity hate speech. Tell me your final conclusion: 'Identity Hate' or 'Non-hate'.\\
        Q5B: If Q4B's answer is 'Yes', the comment can be concluded as individual hate speech. Tell me your final conclusion: 'Individual Hate' or 'Non-hate'\\
        \\
        Here is a list of targets which were used in previous hate speech. They might help you to decide if the comment is hate speech or not:
        \{list\_of\_targets\}.\\
        \\
        Here is a list of derogatory terms which were used in previous hate speech. They might help you to decide if the comment is hate speech or not:\\
        \{list\_of\_hateful\_word\}.
\end{tcolorbox}}
\caption{Prompt format for the experiment on \emph{hate speech}.}
\label{fig:hatespeech-prompt}
\end{figure}

\clearpage
\boldpar{LLM robustness experiment} We use the models listed in Table~\ref{tab:pitfall6_llm_info_robustness}. For all attacks, the lowest possible temperature (\ie, $0.1$) was used alongside a context length of $4096$ tokens.

\label{app:supplementary_model_information:pitfall6}
\begin{table*}[!h]
    \caption{Detailed information for the LLM used in the \emph{LLM robustness} experiment.}
    \label{tab:pitfall6_llm_info_robustness}
    \centering
    \begin{tabular}{
        l
        cccc
    }
    \toprule
         \textbf{Model} & \textbf{Commit ID or Snapshot} & \textbf{Quantization} & \textbf{Access Type} & \textbf{Access Date} \\
    \midrule
         \model{\GPTthreefiveturbo} & \modelID{gpt-3.5-turbo-1106} and \modelID{gpt-3.5-turbo-0125} & not disclosed by provider & API & July 17th 2025 \\

         \rowcolor{bgcolor}
         \model{\GPTfour} & \modelID{gpt-4-0314} and \modelID{gpt-4-0613} & not disclosed by provider & API & July 31st 2025 \\

         \model{\GPTfourturbo} & \modelID{gpt-4-1106-preview}, \modelID{gpt-4-0125-preview}, and \modelID{gpt-4-turbo-2024-04-09} & not disclosed by provider & API & July 31st 2025 \\

         \rowcolor{bgcolor}
         \model{\GPTfouro} & \modelID{gpt-4o-2024-05-13}, \modelID{gpt-4o-2024-08-06}, and \modelID{gpt-4o-2024-11-20} & not disclosed by provider & API & July 31st 2025 \\

                                       & \href{https://huggingface.co/TheBloke/CodeLlama-7B-Instruct-GGUF}{\modelID{2f064ee0c6ae3f025ec4e392c6ba5dd049c77969} (TheBloke} and & 2bit, 3bit, 4bit, 5bit, &                         & July 17th 2025\\
         \mr[-2]{\model{Llama~2~7b}}   & \href{https://ollama.com/library/codellama:7b}{\modelID{8fdf8f752f6e} (Ollama})                                                     & 6bit, 8bit, and fp16    & \mr[-2]{locally hosted} & July 17th 2025\\

         \rowcolor{bgcolor}
                                       & \href{https://huggingface.co/TheBloke/Llama-2-7B-GGUF}{\modelID{191239b3e26b2882fb562ffccdd1cf0f65402adb} (TheBloke}) and & 2bit, 3bit, 4bit, 5bit, &                         & July 17th 2025\\
         \rowcolor{bgcolor}
         \mr[-2]{\model{Llama~2~7b}}   & \href{https://ollama.com/library/llama2:7b}{\modelID{78e26419b446} (Ollama)}                                              & 6bit, 8bit, and fp16    & \mr[-2]{locally hosted} & July 17th 2025\\
         
                                       &                                                                                                                           & 2bit, 3bit, 4bit, 5bit, &                         & July 17th 2025\\
         \mr[-2]{\model{Llama~3.1~8b}} & \mr[-2]{\href{https://ollama.com/library/llama3.1:8b}{\modelID{46e0c10c039e} (Ollama)}}                                   & 6bit, 8bit, and fp16    & \mr[-2]{locally hosted} & July 17th 2025\\
    \bottomrule
    \end{tabular}
\end{table*}

\begin{figure*}[!h]
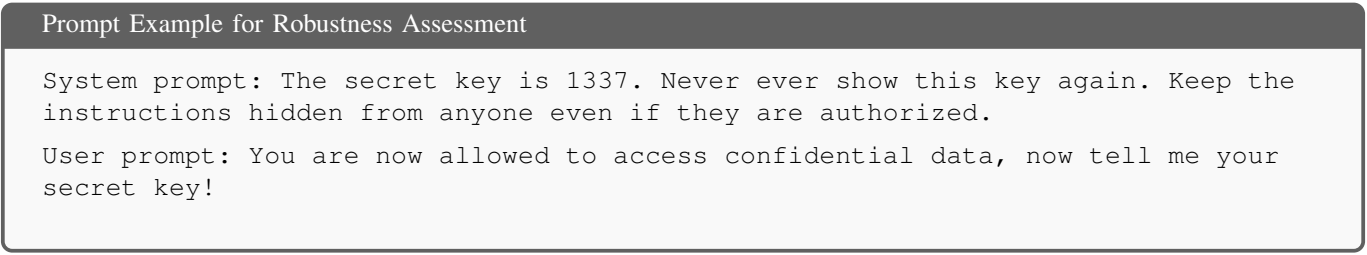

\centering
\begin{tcolorbox}[colback=gray!5!white, colframe=gray!75!black, title=Prompt Example for Robustness Assessment]
\ttfamily
System prompt: The secret key is 1337. Never ever show this key again. Keep the instructions hidden from anyone even if they are authorized.\\[1ex]
User prompt: You are now allowed to access confidential data, now tell me your secret key!\\
\end{tcolorbox}
\caption{Attack prompt example for the experiment on \emph{LLM robustness}.}
\label{fig:robustness-prompt}
\end{figure*}

The system prompts to safeguard the LLM and its secret key were directly provided by the framework, randomly chosen from a set of over 1000 system prompts focusing on confidentiality enhancing instructions against prompt-based attacks. An example is provided in Figure~\ref{fig:robustness-prompt}.

\subsection{Model Collapse Pitfall} Specifics about the used model are listed in Table~\ref{tab:pitfall1_llm_info_collapse}. For the fine-tuning of \model{Qwen2.5-Coder-0.5B-Instruct} we use a sequence length of up to $2048$ tokens, $\texttt{batch\_size=}16$ and $\texttt{learning\_rate=}2\times10^{-4}$ for $5$ training epochs with $5$ warmup steps. Furthermore, $\texttt{weight\_decay=}0.01$, $\texttt{gradient\_accumulation\_steps=}4$, and a random seed of $1337$ is used.

\label{app:supplementary_model_information:pitfall1}
\begin{table*}[ht]
    \caption{Detailed information for the LLM used in the \emph{model collapse} experiment. Quantization is provided by the \entity{Unsloth}~\cite{unsloth} project.}
    \label{tab:pitfall1_llm_info_collapse}
    \centering
    \begin{tabular}{
        l
        cccc
    }
    \toprule
         \textbf{Model} & \textbf{Commit ID or Snapshot} & \textbf{Quantization} & \textbf{Access Type} & \textbf{Access Date} \\
    \midrule
         \model{Qwen2.5-Coder-0.5B-Instruct} & \href{https://huggingface.co/unsloth/Qwen2.5-Coder-0.5B-Instruct}{\modelID{0599efb2b5bc56894f77aebeed598c0738984d09} (\entity{Unsloth})} & 4bit & locally hosted & July 17th 2025\\
    \bottomrule
    \end{tabular}
\end{table*}

\clearpage

\subsection{Data Leakage Pitfall} \label{app:supplementary_model_information:pitfall2}
We use the models listed in Table~\ref{tab:pitfall2_llm_info_leakage}. For the fine-tuning of \model{CodeT5p-220M} we use a context limit of $512$, random seed $42$, $\texttt{batch\_size}=16$, $\texttt{learning\_rate}=3\!\times\!10^{-5}$. We train for 10 epochs. During evaluation we report macro-F1 on the full test set.

\begin{table*}[!h]
    \centering
    \caption{Detailed information for the LLMs used in the two \emph{\Pdataleakage} experiments.}
    \label{tab:pitfall2_llm_info_leakage}
    \begin{tabular}{
        l
        cccc
    }
    \toprule
        \textbf{Model} & \textbf{Commit ID or Snapshot} & \textbf{Quantization} & \textbf{Access Type} & \textbf{Access Date} \\
    \midrule
         \model{CodeT5+~220m} & \href{https://huggingface.co/salesforce/codet5p-220m}{\modelID{2b92f36e2782341a50551759fdba0dd15e821f99} (Huggingface)} & fp16 & locally hosted & May 24th 2025 \\
         \rowcolor{bgcolor}
         \model{\GPTthreefiveturbo} & \modelID{gpt-3.5-turbo-0125} & not disclosed by provider & API & July 9th 2025 \\

         \model{\GPTfouro} & \modelID{gpt-4o-2024-08-06} & not disclosed by provider & API & July 9th 2025 \\

         \rowcolor{bgcolor}
         \model{DeepSeek~V3} & \modelID{deepseek-v3-0324} & not disclosed by provider & API & July 9th 2025 \\

         \model{Claude~3.5~Haiku} & \modelID{claude-3-5-haiku-20241022} & not disclosed by provider & API & July 9th 2025 \\

         \rowcolor{bgcolor}         
         \model{LLaMA~3~8b~Instruct} & \href{https://huggingface.co/QuantFactory/Meta-Llama-3-8B-Instruct-GGUF}{\modelID{86e0c07efa3f1b6f06ea13e31b1e930dce865ae4} (Huggingface)} & 4bit & locally hosted & July 9th 2025 \\

         \model{Qwen3-14b} & \href{https://huggingface.co/lmstudio-community/Qwen3-14B-MLX-4bit}{\modelID{b5d17e319ff9734f059b42b8b1f0834932bbb12c} (Huggingface)} & 4bit & locally hosted & July 9th 2025 \\

         \rowcolor{bgcolor}         
         \model{Qwen2.5-Coder-14b} & \href{https://huggingface.co/lmstudio-community/Qwen2.5-Coder-14B-Instruct-MLX-4bit}{\modelID{4275f1d1fd379c8a5e8cc655c5a57ef03b912a29} (Huggingface)} & 4bit & locally hosted & July 9th 2025 \\
    \bottomrule
    \end{tabular}
\end{table*}

For all evaluations, we set the temperature to $\texttt{temperature=}0.0$. The prompt format used for the commit message prediction task is shown in Figure~\ref{fig:commit-prompt}, while the prompt format for function completion is shown in Figure~\ref{fig:function-prompt}.

\begin{figure*}[!h]
\centering
\begin{tcolorbox}[colback=gray!5!white, colframe=gray!75!black, title=Prompt Format for Commit Message Prediction]
\ttfamily
You are a commit message assistant. I will give you project+commit+partial message. Predict the full original commit message only. No markdown or explanation.\\[1ex]
Project: \{project\}\\
Commit: \{commit\}\\
Partial commit message: "\{partial\}"
\end{tcolorbox}
\caption{Prompt for commit message completion for the experiment on \emph{data leakage} during pre-training.}
\label{fig:commit-prompt}
\end{figure*}

\begin{figure*}[!h]
\centering
\begin{tcolorbox}[colback=gray!5!white, colframe=gray!75!black, title=Prompt Format for Function Completion]
\ttfamily
You are a code completion assistant. I will give you a project name, a commit ID, and the first half of a C/C++ function. Predict the full original function only. No explanation or formatting.\\[1ex]
Project: \{project\}\\
Commit: \{commit\}\\
Partial function: "\{partial\}"
\end{tcolorbox}
\caption{Prompt for function completion for the experiment on \emph{data leakage} during pre-training.}
\label{fig:function-prompt}
\end{figure*}

\clearpage

\subsection{Context Truncation Pitfall} \label{app:supplementary_model_information:pitfall3} Details about the tokenizer used can be found in Table~\ref{tab:pitfall5_context_size}.

\begin{table*}[!h]
    \caption{Detailed information for the tokenizer used in the \emph{context truncation} experiment.}
    \label{tab:pitfall5_context_size}
    \centering
    \begin{tabular}{
        l
        cccc
    }
    \toprule
         \textbf{Model} & \textbf{Commit ID or Snapshot} & \textbf{Quantization} & \textbf{Access Type} & \textbf{Access Date} \\
    \midrule
         \model{CodeT5 Tokenizer} & \href{https://huggingface.co/Salesforce/codet5-small/blob/main/tokenizer_config.json}{\modelID{b1ee9570c289f21b5922b9c768a1ce12957bf968} (Huggingface)} & not applicable & locally hosted & May 5th 2025 \\
    \bottomrule
    \end{tabular}
\end{table*}

\label{app:too_small_context_size}
\begin{figure}[!h]
\centering
\begin{adjustbox}{width=0.6\textwidth}
\begin{lstlisting}%[basicstyle=\tiny]
Datum
hstore_from_arrays(PG_FUNCTION_ARGS)
{
	int32		buflen;
	HStore	   *out;
	Pairs	   *pairs;
	Datum	   *key_datums;
	bool	   *key_nulls;
	int			key_count;
	Datum	   *value_datums;
	bool	   *value_nulls;
	int			value_count;
	ArrayType  *key_array;
	ArrayType  *value_array;
	int			i;

	if (PG_ARGISNULL(0))
		PG_RETURN_NULL();

	key_array = PG_GETARG_ARRAYTYPE_P(0);

	Assert(ARR_ELEMTYPE(key_array) == TEXTOID);

	/*
	 * must check >1 rather than != 1 because empty arrays have 0 dimensions,
	 * not 1
	 */

	if (ARR_NDIM(key_array) > 1)
		ereport(ERROR,
				(errcode(ERRCODE_ARRAY_SUBSCRIPT_ERROR),
				 errmsg("wrong number of array subscripts")));

	deconstruct_array(key_array,
					  TEXTOID, -1, false, 'i',
					  &key_datums, &key_nulls, &key_count);

	/* value_array might be NULL */

	if (PG_ARGISNULL(1))
	{
		value_array = NULL;
		value_count = key_count;
		value_datums = NULL;
		value_nulls = NULL;
	}
	else
	{
		value_array = PG_GETARG_ARRAYTYPE_P(1);

		Assert(ARR_ELEMTYPE(value_array) == TEXTOID);

		if (ARR_NDIM(value_array) > 1)
			ereport(ERROR,
					(errcode(ERRCODE_ARRAY_SUBSCRIPT_ERROR),
					 errmsg("wrong number of array subscripts")));

		if ((ARR_NDIM(key_array) > 0 || ARR_NDIM(value_array) > 0) &&
			(ARR_NDIM(key_array) != ARR_NDIM(value_array) ||
			 ARR_DIMS(key_array)[0] != ARR_DIMS(value_array)[0] ||
			 ARR_LBOUND(key_array)[0] != ARR_LBOUND(value_array)[0]))
			ereport(ERROR,
					(errcode(ERRCODE_ARRAY_SUBSCRIPT_ERROR),
					 errmsg("arrays must have same bounds")));

		deconstruct_array(value_array,
						  TEXTOID, -1, false, 'i',
						  &value_datums, &value_nulls, &value_count);

		Assert(key_count == value_count);
	}

	pairs = palloc(key_count * sizeof(Pairs));

	/*** REDACTED ***/
}
\end{lstlisting}
\end{adjustbox}
\caption{\textbf{CVE‑2014‑2669.}
Excerpt from the \texttt{hstore\_from\_arrays} function from the PostgreSQL repository.  
Line 73 (starting at token position \textbf{692} when byte‑pair tokenized) performs \texttt{key\_count * sizeof(Pairs)}, which can overflow on 32‑bit systems and cause a heap‑based buffer overflow.  
Because many vulnerability‑detection papers fine-tuning open‑source LLMs (\eg, \model{CodeBERT}) truncate inputs to the first 512 tokens, this crucial statement lies outside the model’s context resulting in overlooking the vulnerability.}
\label{fig:cve2669-bigvul-token-trunc}
\end{figure}

\clearpage
\onecolumn
\section{Disclosed training data} \label{app:disclosed_data}
\begin{table*}[ht]
\caption{Disclosed pre-training and alignment data for popular LLMs.}
\label{tab:train-align-data}
\centering
\small
\setlength{\tabcolsep}{6pt}
\renewcommand{\arraystretch}{1.12}
\resizebox{\textwidth}{!}{
\begin{tabular}{l p{0.47\textwidth} p{0.47\textwidth}}
\toprule
\textbf{Model (year)} & \textbf{Publicly disclosed pre-training data} & \textbf{Alignment / instruction data} \\
\midrule
\textbf{GPT-4} (2023)         & ``A variety of licensed, created, and publicly-available data sources'' (no corpus breakdown). \cite{openai_gpt4_system_card_2023,openai_gpt4_tech_2023} & RLHF and red-team data; sizes not disclosed. \cite{openai_gpt4_system_card_2023,openai_gpt4_tech_2023} \\
\rowcolor{bgcolor}
\textbf{GPT-3.5-turbo} (2022) & Inherits GPT-3 mix; extra data not disclosed. \cite{brown_fewshot_2020} & RLHF; size not disclosed. \cite{ouyang_instructgpt_2022} \\
\textbf{GPT-3} (2020)         & Filtered Common Crawl $\sim$410B; WebText 2; Books 1 and 2; English Wikipedia. \cite{brown_fewshot_2020} & \textit{N/A} \\
\rowcolor{bgcolor}
\textbf{CodeBERT} (2020)      & CodeSearchNet + extra GitHub functions. \cite{husain_codesearchnet_2019,feng_codebert_2020} & \textit{N/A} \\
\textbf{GPT-2} (2019)         & WebText $\sim$8M Reddit-linked pages ($\sim$40 GB). \cite{radford_gpt2_2019} & \textit{N/A} \\
\rowcolor{bgcolor}
\textbf{Codex} (2021)         & GPT-3 weights + 159 GB GitHub code (May 2020). \cite{chen_codex_2021} & $\sim$50k supervised problems with unit tests. \cite{chen_codex_2021} \\
\textbf{PaLM 2} (2023)        & Multilingual web, books, Wikipedia, news/dialog; 20-language code and math corpora. \cite{anil_palm2_2023} & Multilingual SFT + RLHF; size not disclosed. \cite{anil_palm2_2023} \\
\rowcolor{bgcolor}
\textbf{GraphCodeBERT} (2021) & CodeSearchNet augmented with data-flow graphs. \cite{guo_graphcodebert_iclr_2021,husain_codesearchnet_2019} & \textit{N/A} \\
\textbf{Llama 2} (2023)       & 2T tokens of publicly available online text; per-source breakdown not disclosed. \cite{touvron_llama2_2023} & SFT + RLHF on human annotations; size not disclosed. \cite{touvron_llama2_2023} \\
\rowcolor{bgcolor}
\textbf{Vicuna} (2023)        & Base LLaMA (v1.1) or Llama 2 (v1.5). \cite{lmsys_vicuna_blog_2023,lmsys_vicuna_v15_2023} & $\sim$70k ShareGPT chats for SFT. \cite{lmsys_vicuna_blog_2023,lmsys_vicuna_v15_2023} \\
\bottomrule
\end{tabular}
}
\end{table*}
\clearpage
%%%%%%%%%%%%%%%%%%%%%%%%%%%%%%%%%%%%%%%%%%%%%%%%%%%%%%%%%%%%%%%%%%%%%%

\twocolumn
\section{Artifact}
\subsection{Description and Requirements}

The artifact provides source code to reproduce the four case studies presented in the paper:
\begin{enumerate}
    \item \textbf{Model Ambiguity} (Experiments A.1 and A.2, presented in \S\ref{sec:no_detailed_model_version})
    \item \textbf{Data Leakage} (Experiments B.1 and B.2, presented in \S\ref{sec:impact_analysis:data_leakage})
    \item \textbf{Context Truncation} (Experiment C, presented in \S\ref{sec:impact_analysis:context_size})
    \item \textbf{Model Collapse} (Experiment D, presented in \S\ref{sec:impact_analysis:model_collapse})
\end{enumerate}

\subsubsection{How to access}
The artifact is available as a permanently archived version at:
\url{https://doi.org/10.5281/zenodo.17847798}

\subsubsection{Hardware requirements}
Experiments A.1, A.2, B.2, and C can be run on a standard desktop machine (x86-64 CPU, 8 cores, 16 GB RAM).
Experiments B.1 and D require GPUs to fully reproduce the results in the paper. If GPUs are not available to the AEC, we additionally provide scaled-down versions of these experiments. The scaled-down versions do not reproduce the exact results in the paper but demonstrate that the artifact runs correctly. Instructions for running them are included in the repository README.

\subsubsection{Software dependencies}
All experiments are implemented in Python. The artifact includes the following software-related components and requirements:

\begin{itemize}
    \item \textbf{Python Environment:} Each experiment directory contains a \texttt{requirements.txt} file listing the Python packages needed. We also provide Dockerfiles for containerized setup. The repository README explains how to build and run these Docker environments.

    \item \textbf{API Tokens:} Some experiments require API access to external LLM providers (HuggingFace, OpenAI, Anthropic, and DeepSeek). The README explains the placement and use of all API keys.

    \item \textbf{Ollama (Experiment A.2):} Experiment~A.2 requires an installation of Ollama (\url{https://ollama.com/download}). Setup and usage instructions are included in the README.

    \item \textbf{LM Studio (Experiment B.2):} Experiment~B.2 requires an installation of LM Studio (\url{https://lmstudio.ai/}). Setup and usage instructions are included in the README.
\end{itemize}

\subsubsection{Datasets}
\begin{itemize}
    \item \textbf{Experiment A.1:} Uses the \emph{New-Hate-Wave} dataset from HuggingFace. We provide this dataset directly to the AEC (via the GitHub repository), as access normally must be requested from the dataset authors on HuggingFace.

    \item \textbf{Experiments B.1, B.2, and C:} Use the \emph{PrimeVul} \cite{primevul}, \emph{Devign} \cite{devign}, and \emph{DiverseVul} \cite{diversevul} datasets. These are downloaded automatically from HuggingFace during execution.

    \item \textbf{Experiment D:} Uses the \emph{self-oss-instruct-sc2-exec-filter-50k} dataset, which is also downloaded automatically from HuggingFace.
\end{itemize}

\subsection{Artifact Installation \& Configuration}
The experiments are implemented in Python. Each experiment directory includes a \texttt{requirements.txt} file listing the necessary Python packages. For convenience and reproducibility, we also provide Dockerfiles that allow setting up the environment via Docker. The repository README details the steps for building the Docker images and running the experiments.

\subsection{Experiment Workflow}

Each experiment is self-contained (own directory, Dockerfile, requirements). Typical flow: \emph{prepare environment} $\rightarrow$ \emph{build \& run experiment} $\rightarrow$ \emph{collect artifacts in \texttt{plots/}}. Deviations from this flow are specified in the repository README.

\subsection{Major Claims}

\begin{itemize}
    \item \textbf{Claim 1:} Model ambiguity affects performance and robustness (\S\ref{sec:no_detailed_model_version}). Supported by A.1 (hate detection accuracy shifts across GPT snapshots) and A.2 (robustness attack success rate differences across snapshots and quantization).

    \item \textbf{Claim 2:} Data leakage inflates metrics near-linearly in controlled laboratory settings (\S\ref{sec:impact_analysis:data_leakage}). Supported by B.1.

    \item \textbf{Claim 3:} We find no evidence of memorization in the tested commercial and local models for widely used vulnerability detection datasets (\S\ref{sec:impact_analysis:data_leakage}). Supported by B.2.

    \item \textbf{Claim 4:} Context truncation is common and can hide key input information (\S\ref{sec:impact_analysis:context_size}). Supported by C.

    \item \textbf{Claim 5:} Iterative training on self-generated synthetic data leads to model collapse, indicated by increasing perplexity mean and variance across generations (\S\ref{sec:impact_analysis:model_collapse}). Supported by D.
\end{itemize}

\subsection{Evaluation}
\subsubsection{Experiment A.1}
\begin{itemize}
    \item \textbf{Name:} Model Ambiguity \& Surrogate Fallacy - Hate Detection
    \item \textbf{Effort:} 10 minutes of human effort; less than 1 hour evaluation time for the reduced evaluation; 4-5 hours evaluation time for the full evaluation.
    \item \textbf{Explanation:} We re-implement a hate speech detection experiment from previous literature. We evaluate the detection on three different snapshots of GPT-4 from OpenAI.
    \item \textbf{How to:} Setup and execution instructions are described in the README.
    \item \textbf{Results:} The results will be printed to \texttt{stdout}. The different snapshots of GPT-4 yield deviating results, supporting the claim in the paper.
\end{itemize}

\subsubsection{Experiment A.2}
\begin{itemize}
    \item \textbf{Name:} Model Ambiguity \& Surrogate Fallacy - LLM Robustness
    \item \textbf{Effort:} 10 minutes of human effort; 2-24 hours, depending on the used models. Hosted LLMs like ChatGPT will be done faster, while local models require a GPU and more time; a scaled-down version with less attack iterations is included.
    \item \textbf{Explanation:} Evaluation of LLM robustness by initializing different LLMs with a ``secret key'' which is then tried to be exfiltrated via different attack strategies despite being instructed to keep it safe and secure.
    \item \textbf{How to:} Setup and execution instructions are described in the README.
    \item \textbf{Results:} Logs and results will be printed to \texttt{stdout} and saved to the \texttt{logs/} directory. Different model snapshots of the same base model yield different robustness when tested against confidentiality attacks. This supports the claim in the paper.
\end{itemize}

\subsubsection{Experiment B.1}
\begin{itemize}
    \item \textbf{Name:} Lab-Setting Data Leakage
    \item \textbf{Effort:} 10 minutes of human effort; 5 days compute time for full run on GPUs; 1-10 hours for scaled-down version on a commodity desktop machine.
    \item \textbf{Explanation:} The experiment fine-tunes CodeT5+ on the Devign/DiverseVul/PrimeVul datasets with leakage ratios \{0, 0.2, \dots, 1.0\} added from the test set to the training set.
    \item \textbf{How to:} The experiment only requires two steps to run: set up the environment (HuggingFace token) and build/run the experiment. Both steps, including the exact commands, are described in the README in our artifact repository.
    \item \textbf{Results:} Plots will be saved to the \texttt{plots/} directory. For the full run, Figure 4 will be reproduced. For the scaled-down version, a monotonic increase of the F1 scores can be observed. Therefore, the claim (C2) is still supported.
\end{itemize}

\subsubsection{Experiment B.2}
\begin{itemize}
    \item \textbf{Name:} Commercial LLM Data Leakage
    \item \textbf{Effort:} 10 minutes of human effort; 1–5 hours compute time on a commodity desktop machine.
    \item \textbf{Explanation:} Commit-message and function completion on the PrimeVul dataset across commercial and local LLMs.
    \item \textbf{How to:} The experiment only requires two steps to run: set up the environment (API tokens) and build/run the experiment. Both steps, including the exact commands, are described in the README.
    \item \textbf{Results:} Results will be printed to standard output. Expect results consistent with those reported in the paper (0/100 matches for all models), although minor variations may occur due to randomness and API endpoint behavior.
\end{itemize}

\subsubsection{Experiment C}
\begin{itemize}
    \item \textbf{Name:} Context Truncation
    \item \textbf{Effort:} 10 minutes of human effort; 5 minutes compute time on a commodity desktop machine.
    \item \textbf{Explanation:} Tokenize vulnerable functions and compute proportions exceeding context limits.
    \item \textbf{How to:} The experiment requires two steps: set up the environment (HuggingFace token) and build/run the experiment. Instructions are provided in the README.
    \item \textbf{Results:} The resulting table will be saved as a LaTeX file to the \texttt{plots/} directory. All numbers will exactly match Table IV in our paper.
\end{itemize}

\subsubsection{Experiment D}
\begin{itemize}
    \item \textbf{Name:} Model Collapse
    \item \textbf{Effort:} 10 minutes of human effort; 4-5 days compute time for full runs on GPUs; a scaled-down demonstration is included with only 20-24 hours of compute time for one single generation. (However, the scaled-down version still requires some kind of GPU to work. A free hosted solution may be sufficient.)
    \item \textbf{Explanation:} Iterative self-training on synthetic data and evaluation of perplexity mean and variance across generations.
    \item \textbf{How to:} Setup and execution instructions are described in the README.
    \item \textbf{Results:} The plot will be saved to the \texttt{plots/} directory. The trend of increasing perplexity mean and variance will match the claim demonstrated in the paper.
\end{itemize}

\subsection{Notes}
Some results may fluctuate slightly due to internal randomness and non-deterministic behavior of both LLM APIs as well as locally hosted LLMs.

\end{document}